\documentclass[11pt]{iopart}
\usepackage[utf8]{inputenc}
\usepackage{subfiles}
\usepackage[margin=1.in]{geometry}
\usepackage{pgf}  
\usepackage{graphicx,wrapfig}
\usepackage{float}
\usepackage{subcaption}
\usepackage{multicol}
\usepackage{gensymb}
\usepackage[utf8]{inputenc}
\usepackage[english]{babel}
\usepackage{tabu}
\usepackage[title]{appendix}
\usepackage{subcaption}
\usepackage{listings}

\begin{document}

\title{Simulating the Impact of Baffling on Divertor Performance Using SOLPS-ITER}
\date{\vspace{-5ex}}

\author{C Cowley$^{1,2}$, D Moulton$^{2}$, B Lipschultz$^{1}$}

\address{$^{1}$ York Plasma Institute, University of York, Heslington, York, YO10 5DQ, UK}

\address{$^{2}$ UKAEA-CCFE, Culham Science Centre, Abingdon, OX14 3DB, UK}

\ead{cyd@digilab.co.uk}

\maketitle

\begin{abstract}

Strong divertor baffling is a feature expected to have a number of advantages for core-edge integration in tokamaks, yet one which requires much more detailed and extensive study. In this work, the impacts of baffling on a hydrogenic plasma are studied in isolation, through artificial fixed-fraction impurity SOLPS-ITER simulations. These simulations are of the connected double null Super-X divertor on the MAST-U tokamak, with extreme cases of a closed and open divertor. Simulations show a divertor with a tightly baffled entrance can lead to better access to detachment on the outer targets, likely caused by reduced convection upstream. Adding tight baffling at the throat also leads to two orders of magnitude increase in neutral compression, and an order of magnitude more efficient pumping, with more peaked core density profiles. Finally, in contrast to the localised radiation of the closed divertor, the open divertor shows radiation along the entire plasma edge which moves upstream and inward into the core as detachment evolves; leading to sub-10eV temperatures in the near SOL at x-point.

\end{abstract}

\section{Introduction}

If tokamaks are to become energy-producing power-plants, they must find some means of mitigating the intense heat and particle loads expected on surrounding material \cite{zohm2013assessment}. If unmitigated, these loads can damage and quickly erode even the strongest candidate materials for plasma facing components  \cite{loarte2007power,KUANGDIVERTOR}. Reducing these material loads whilst maintaining good core performance is known as the core-edge integration challenge for reactor tokamaks. It is a challenge that has many potential solutions; one of which is operating with an alternative divertor.

Alternative divertors are divertors which leverage some sort of geometric characteristic alternative to the standard single null divertor configuration which is implemented in devices such as ITER \cite{pitts2019physics}. These characteristics include magnetic features, such as the expansion of magnetic flux tube volume through a reduction in poloidal and toroidal fields, known as poloidal and toroidal flux expansion \cite{lipschultz2016sensitivity,Cowley_2022}. Such magnetic features are leveraged by alternative divertors such as the X-divertor, Super-X divertor \cite{valanju2009super}, or Snowflake divertor \cite{ryutov2007geometrical}. 

In addition to magnetic features, alternative divertor design also leverages the physical design of the plasma facing components, and in particular the baffling of the divertor. Implementing a wall structure which closes off the divertor chamber from the main chamber is thought to keep recycled neutrals more confined in the divertor. These so-called closed divertor configurations are predicted to have several benefits, including more efficient pumping and increased power losses \cite{Stangeby_2017}. Such configurations have been implemented in machines such as MAST-U and TCV \cite{reimerdes2021initial,wensing2019solps,havlivckova2015solps}. Additionally, configurations known as tightly baffled divertors, in which material closely surrounds the entire divertor plasma leg, are thought to provide additional benefits for detachment control \cite{umansky2017attainment,umansky2019study}. Such tightly baffled divertors are considered for for devices such as TCV \cite{Sun_2023} and ARC \cite{wigram2019performance}.

In this work we use SOLPS-ITER to simulate the impacts of varying baffling on divertor performance. Though the impact of baffling has been studied previously in simulations \cite{havlivckova2015solps,fil2020separating}, these works focused only on single simulations, and do not include density scans and extrapolation to high power. What's more, previous studies have typically been motivated by experimental prediction or comparison \cite{Sang_2017,Stangeby_2017}, rather than fundamental code experiments. The study presented here is an idealised one which aims to more fundamentally understand the impacts of extreme baffling on a hydrogenic plasma, neglecting the effects of impurity transport. This extreme baffling includes strong baffling where structures ensure a low chance of recycled neutrals escaping the divertor chamber, and open baffling where no structures prohibit the escape of recycled neutrals. A particular focus of this study is how baffling affects detachment, a likely operating regime for future reactors characterised by significant plasma power and pressure losses. Over the following sections, the simulated impact of baffling on detachment access, power balance, pressure loss, fuelling, and detachment evolution will be explored in SOLPS-ITER simulations of MAST-U.

\section{Simulation Setup}

\begin{figure}
    \centering
    \includegraphics[width=0.65\linewidth]{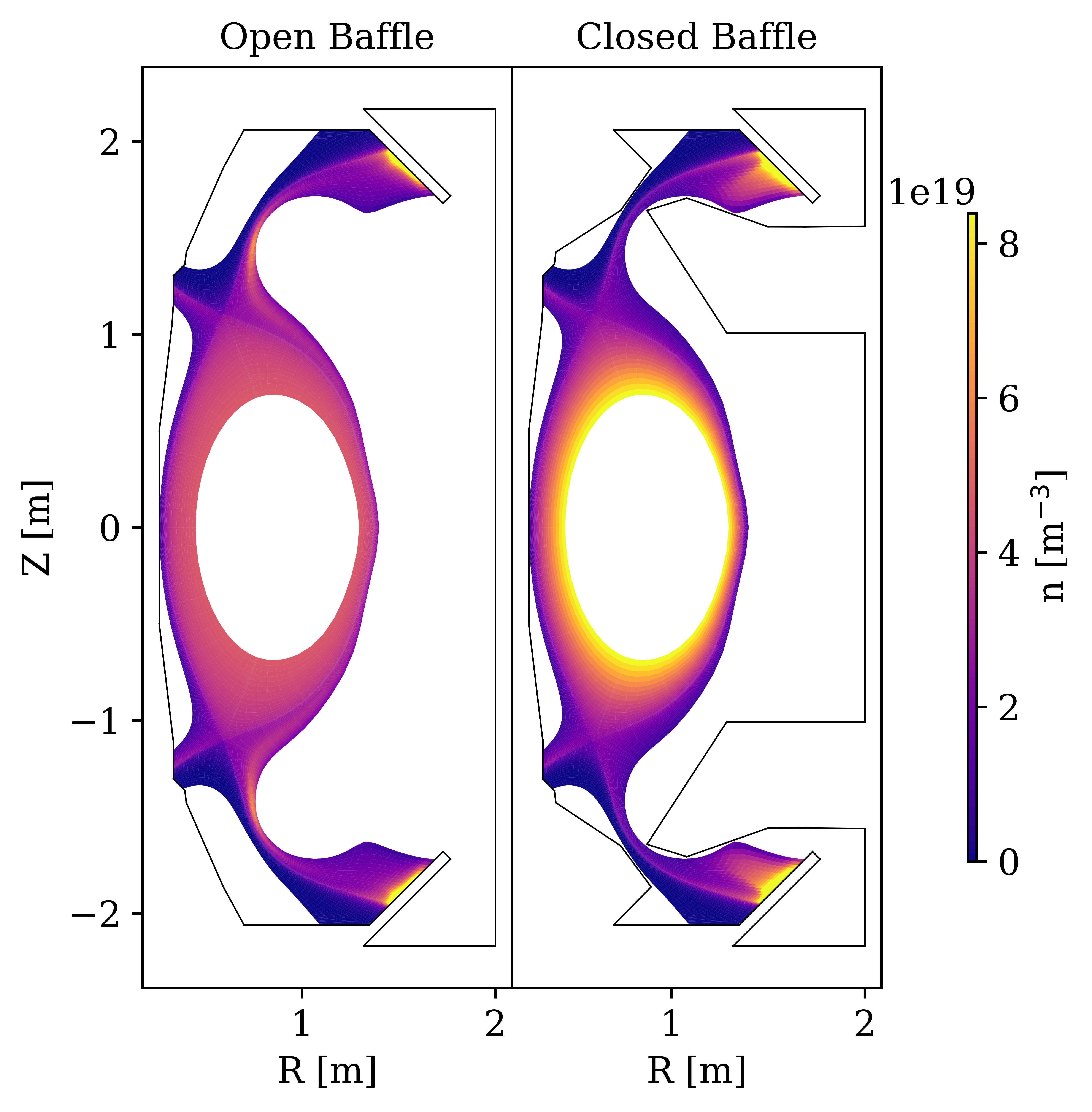}
    \caption{A 2D profile of electron density for an open (left) and closed baffled (right) MAST-U Super-X divertor, at similar target conditions with 12 MW of input power.}
    \label{fig:12MW_ne_Thresh}
\end{figure}

To investigate the isolated impacts of closure and baffling on divertor performance, the code SOLPS-ITER has been used \cite{bonnin2016presentation, SOLPS-ITER}. SOLPS-ITER is a widely used Scrape-Off Layer (SOL) simulator, that can model neutral species kinetically with the Monte Carlo transport code EIRENE \cite{SOLPS-ITER}. Because of this, neutral transport can be effectively modelled throughout the entire chamber, allowing for a realistic study into baffling. The simulations used in this study are of a symmetric double-null Super-X geometry in the MAST-U tokamak \cite{morris2018mast}, shown in Figure \ref{fig:12MW_ne_Thresh}. Neutral-neutral collisions are enabled in these simulations, and the simulations performed are without drifts.

Contrary to the realistic MAST-U walls, these simulations have been performed with two cases of extreme baffling. The first, labelled the `Open' geometry, is effectively one large main chamber, with no material baffling separating the divertor region and core. The second, labelled the `Closed' geometry, has divertor baffling which tightly surrounds the divertor plasma near the x-point, preventing recycled neutrals from escaping the divertor. All plasma facing components are tungsten with 100\% recycling, apart from a vertically symmetric pumping system in the outer divertors with a 1\% pumping fraction. This corresponds to a pumping speed of $\approx$ 25 m$^{3}$s$^{-1}$.

A particularly important aspect of this study is detachment access and control. To facilitate detachment - particularly at higher powers - an artificial nitrogen impurity has been implemented in these simulations, with an impurity fraction of 3\%. The electron cooling function used is shown in \ref{appendix}, and is an analytical approximation for nitrogen at $n_{e} \tau_{\alpha} = 10^{20}$ m$^{-3}$ms (though baffling may change the impurity residence time in reality) \cite{kallenbach2013impurity}. Scans in core density have been performed for each geometry. The conditions range from attached to deeply detached, and roughly 40 simulations have been performed in total.

Another vital aspect of this study is the relevance of baffling as research transitions from present day devices to reactor-like tokamaks. As a consequence, the simulations have been run at three levels of power: 3 MW, 6 MW, and 12 MW, with the 12 MW cases corresponding to $\approx$ 250 MWm$^{-2}$ peak parallel heat fluxes entering the outer divertors. It is important to note that even these extremely high heat fluxes are below what is expected in machines such as STEP or SPARC \cite{hudoba2023divertor,KUANGDIVERTOR}, but running at even higher powers with only artificial nitrogen led to issues accessing detachment in preliminary simulations.

In terms of fuelling, a fixed density boundary condition has been used at the core, with no external puff. This decision was made to simplify the simulations and replicate fuelling by pellets that may be dominant for reactor-like plasmas. Core fuelling ranged from $\approx$ 2$\times$ 10$^{21}$ s$^{-1}$ for attached 3 MW simulations to 2 $\times$ 10$^{23}$ s$^{-1}$ for detached 12 MW simulations.

Radial profiles of anomalous transport coefficients are identical for all simulations presented. Transport coefficients were chosen to replicate low power experimental campaigns on the MAST-U, with particle diffusivity D = 4 m$^{2}$s$^{-1}$ in the core and 2 m$^{2}$s$^{-1}$ in the SOL, and heat diffusivity $\chi$ = 15 m$^{2}$s$^{-1}$ in the core and 7 m$^{2}$s$^{-1}$ in the SOL for both ions and electrons. These transport coefficients led to heat flux widths of $\lambda_{q} \approx 5$ mm mapped to the midplane.

\section{Detachment Access}\label{sec:DetachAcess}

Since some form of detachment is an attractive (and likely necessary) operating regime for reactor-like tokamaks, understanding the impact of baffling on detachment access is vital. To investigate this, the threshold of detachment has been determined for the open and closed MAST-U geometries at all three power levels. This is done by first identifying the location of the detachment front for each simulation, $s_{f,pol}$. Though there are many definitions of the location of a detachment front \cite{Cowley_2022}, here the location is defined as the point at which $T_{e}=5$ eV along the `killer' SOL ring (poloidal flux surface) in the lower outer divertor. The killer SOL ring is the SOL ring in which the attached target heat load on the outer targets peaks - which for these simulations is the third ring outward from the separatrix.

\begin{figure}[ht]
\begin{subfigure}{.32\textwidth}
  \centering
  \includegraphics[width=\linewidth]{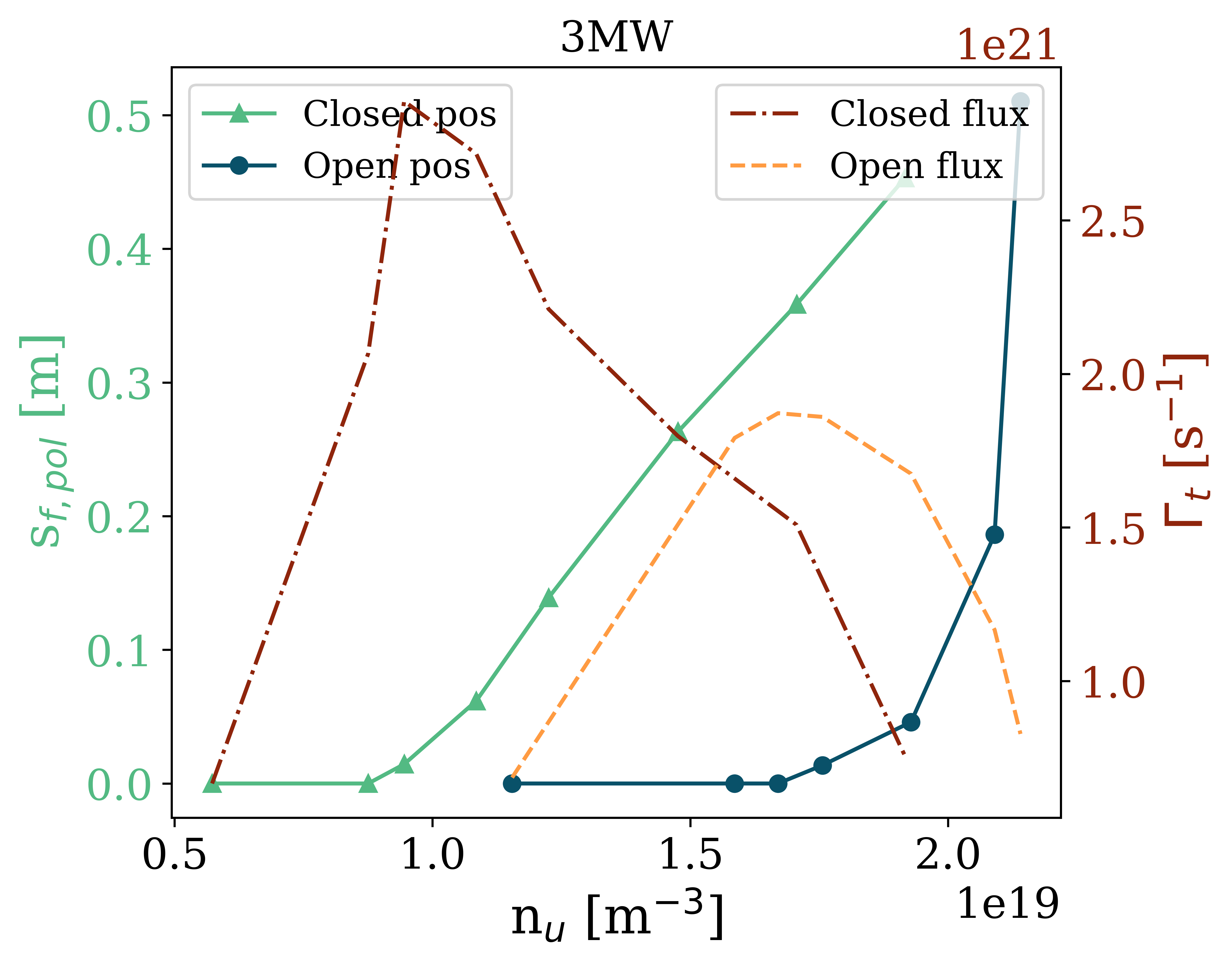}  
\caption{}
\label{fig:BaffleEvolution3MW}
\end{subfigure}
\begin{subfigure}{.32\textwidth}
  \centering
  \includegraphics[width=\linewidth]{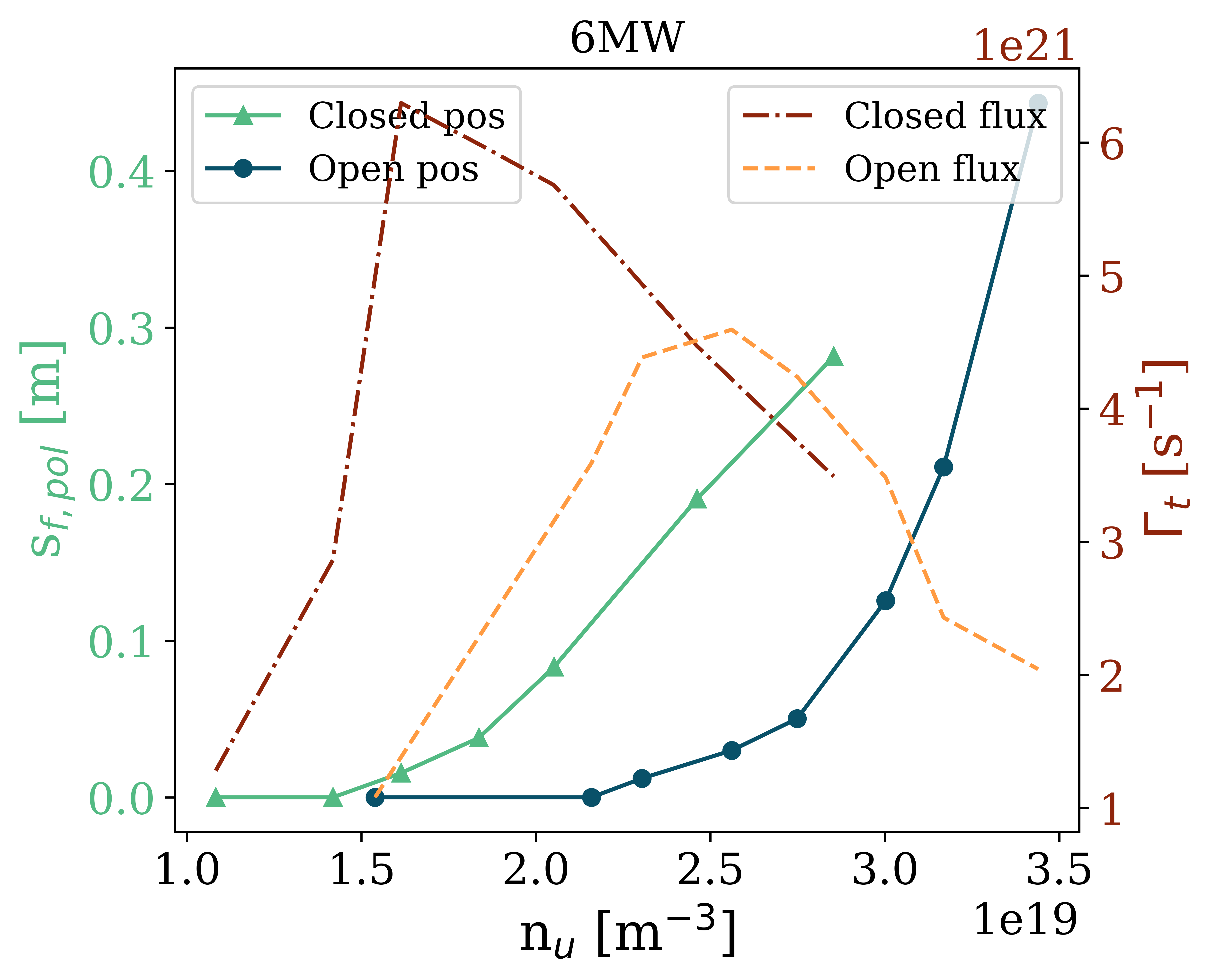} 
\caption{}
\label{fig:BaffleEvolution6MW}
\end{subfigure}
\begin{subfigure}{.32\textwidth}
  \centering
  \includegraphics[width=\linewidth]{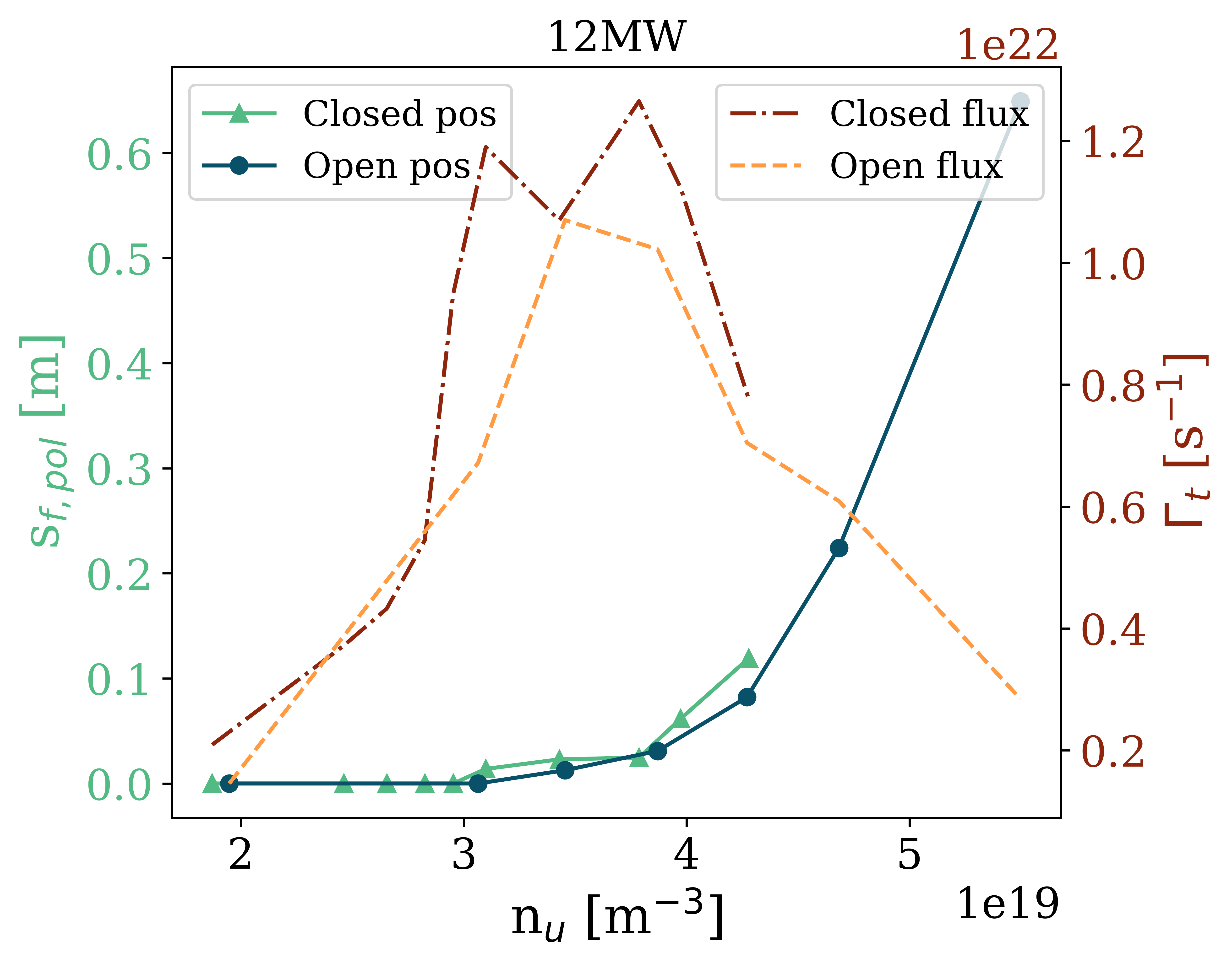} 
\caption{}
\label{fig:BaffleEvolution12MW}
\end{subfigure}
\caption{The lower outer detachment front location, plotted as a function of outer midplane density on the killer flux tube for SOLPS-ITER simulations of closed and open MAST-U geometries, at a) 3 MW, b) 6 MW, and c) 12 MW of input power. Also plotted is the ion particle flux on the killer flux tube at the target.}
\label{fig:BaffleEvolution}
\end{figure}

For each density scan the detachment front position on the lower-outer divertor is plotted as a function of the midplane electron density on the killer flux tube, $n_{u}$, in Figure \ref{fig:BaffleEvolution}. From these simulations, the detachment threshold can be determined as the density corresponding to the first simulation with a detachment front off the target (or in other words, a target temperature below 5 eV). Another common definition of the threshold of detachment is the rollover of target ion flux. As such, Figure \ref{fig:BaffleEvolution} also shows the ion target particle flux on the `killer` flux tube as a function of midplane density.

\begin{figure}[ht]
\begin{subfigure}{.32\textwidth}
  \centering
  \includegraphics[width=\linewidth]{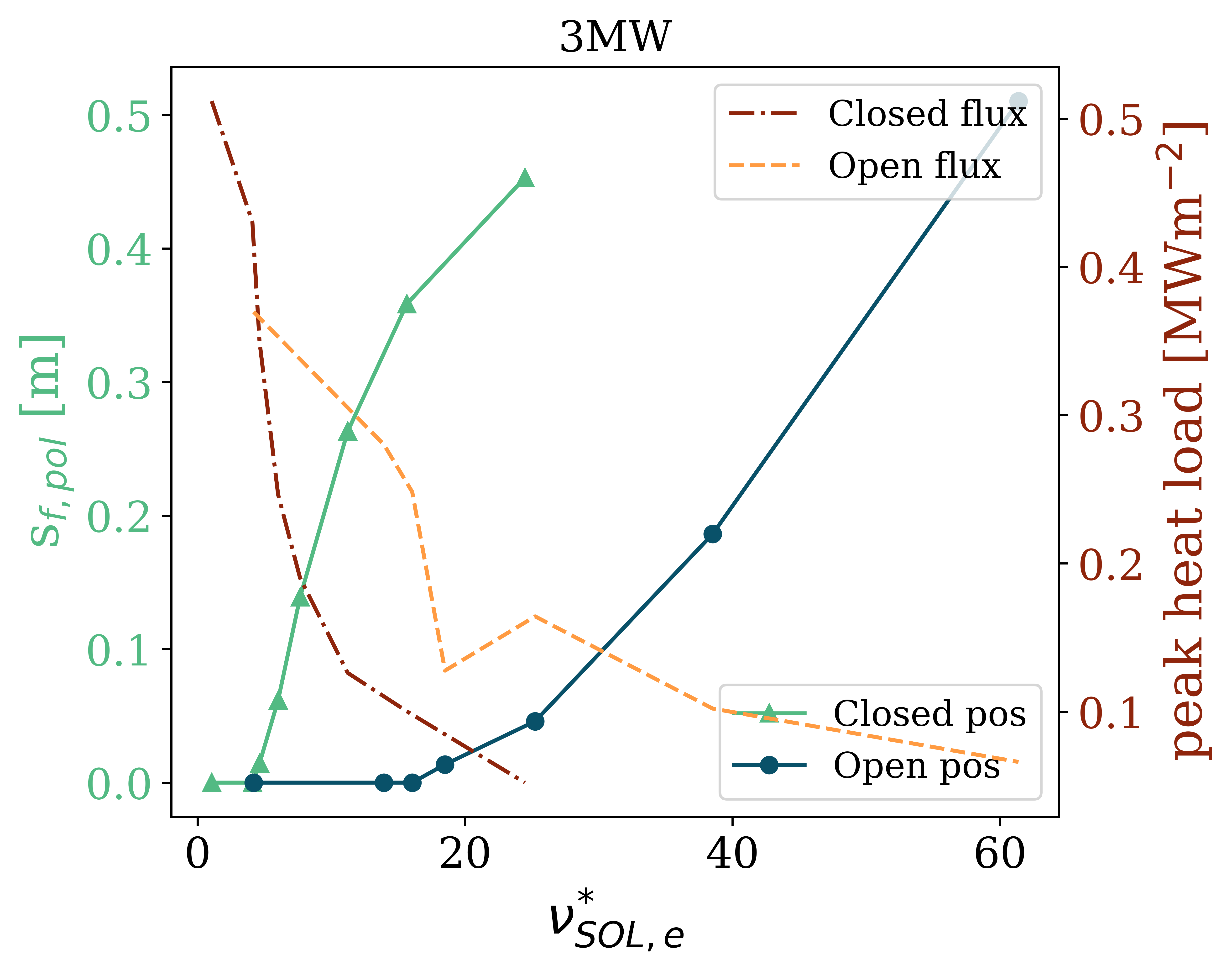}  
\caption{}
\label{fig:BaffleFluxEvolution3MW}
\end{subfigure}
\begin{subfigure}{.32\textwidth}
  \centering
  \includegraphics[width=\linewidth]{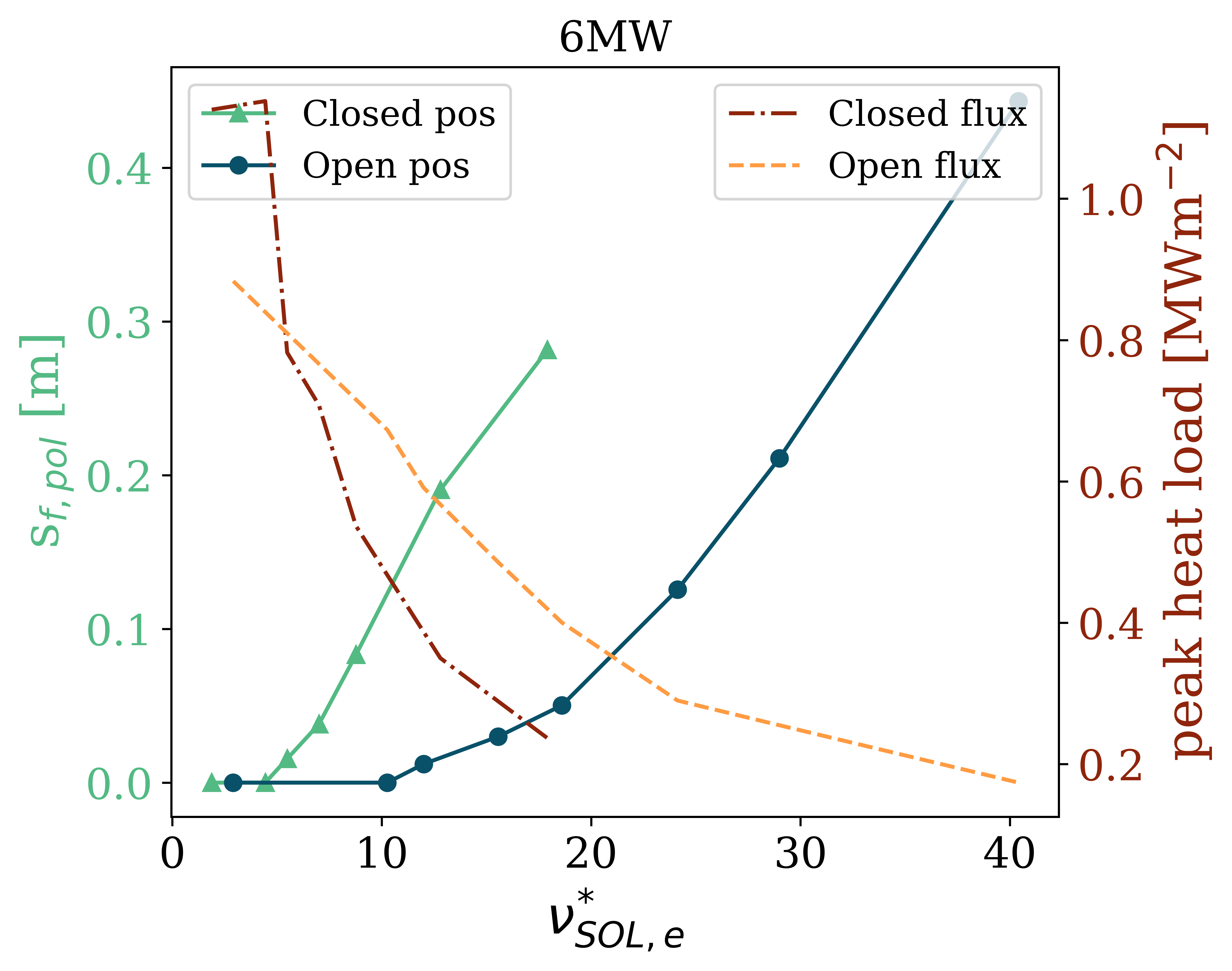} 
\caption{}
\label{fig:BaffleFluxEvolution6MW}
\end{subfigure}
\begin{subfigure}{.32\textwidth}
  \centering
  \includegraphics[width=\linewidth]{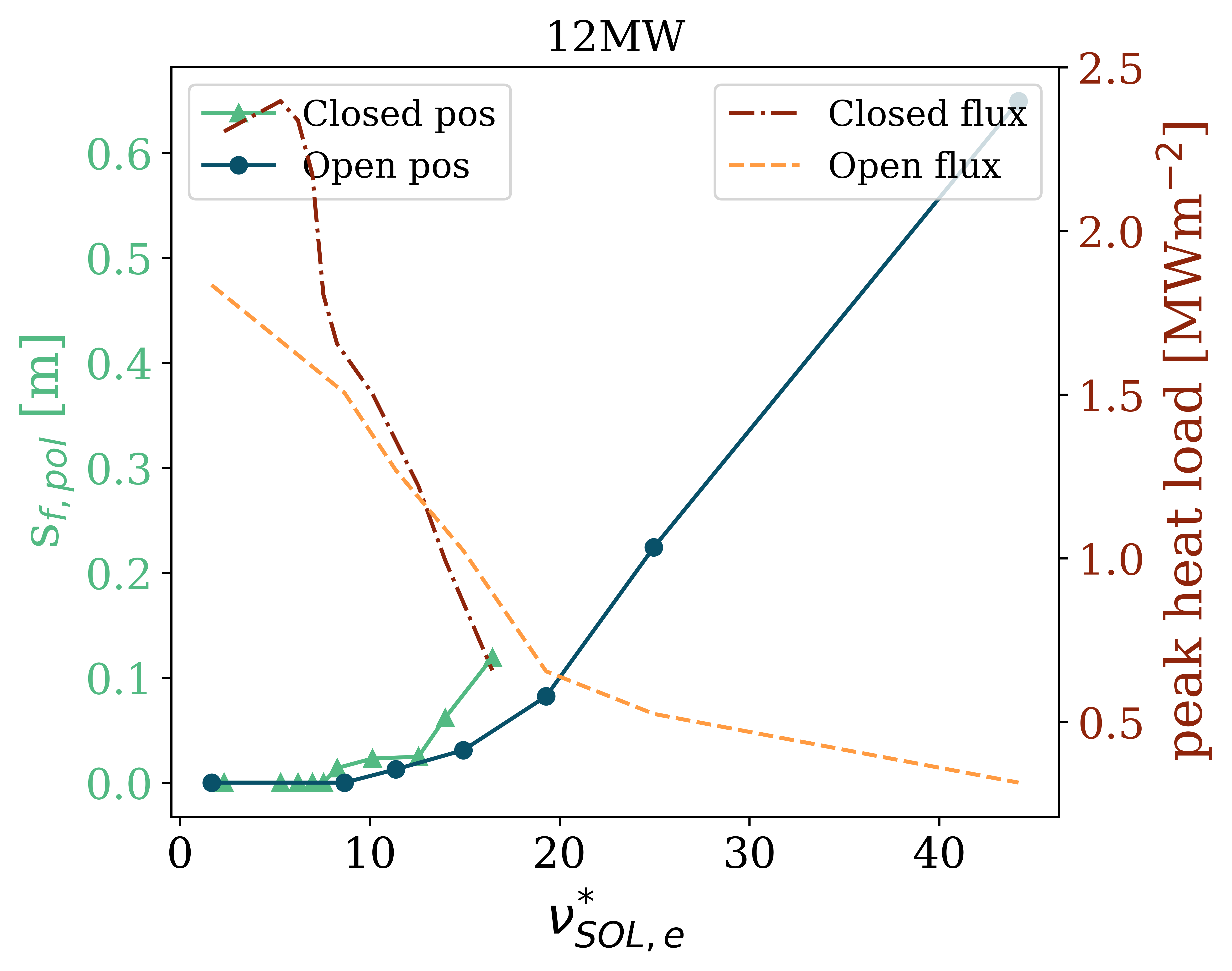} 
\caption{}
\label{fig:BaffleFluxEvolution12MW}
\end{subfigure}
\caption{The lower outer detachment front location, plotted as a function of outer midplane electron collisionality on the killer flux tube for SOLPS-ITER simulations of closed and open MAST-U geometries, at a) 3 MW, b) 6 MW, and c) 12 MW of input power. Also plotted is the peak heat load on the outer divertor targets.}
\label{fig:BaffleFluxEvolution}
\end{figure}

In Figure \ref{fig:BaffleEvolution}, it is clear to see the closed geometry has easier detachment access, since the upstream density corresponding to the first simulation with $s_{f,pol}>$0 is lower. In fact, at 3 MW, the open case detaches at a 91\% higher upstream density than the closed geometry. This difference indicates baffling can have a strong impact on detachment access, which is consistent with previous studies and may help to explain differing detachment results across experiments or simulations with varying baffling \cite{fil2020separating}. The closed geometry also appears to have less sensitive detachment front movement with respect to changes in density. This corresponds to a higher window of detachment, and is discussed further in Section \ref{sec:evolution}. At high power, however, this difference reduces: at 6 MW the difference in density at detachment onset is 52\%; at 12 MW the difference is only 8\%.  

Figure \ref{fig:BaffleEvolution} also shows the point at which the 5 eV front detaches from the target corresponds well to the rollover in ion target flux on the killer flux tube. However, when the detachment thresholds are significantly different, the peak in target fluxes are also different, with the closed geometry displaying higher peaks. Hence, if the primary goal of detachment is to reduce ion target fluxes, the closed case is not necessarily better than the open case. Though the closed geometry may still be better in terms of sputtering rates, since it is able to access lower temperatures more easily.

The density required for detachment access, compared above for the two divertors, is one measure of divertor performance. In addition to this, it is important to investigate how the peak target heat loads vary with upstream conditions. In particular, the upstream collisionality is an important parameter, since a low upstream collisionality has been shown to be beneficial for current drive \cite{poli2012electron} and pedestal pressures and stability, in addition to enhancing density peaking \cite{Dunne_2017,Frassinetti_2017,Frassinetti_2021}. Consequently, achieving a cold divertor with the lowest upstream collisionality is of great importance for core-edge integration. 

Figure \ref{fig:BaffleFluxEvolution} shows the variation in peak target heat load (this is the total heat load, including surface recombination) with upstream electron collisionality $\nu^{*}_{SOL,e} \approx 10^{-16} \frac{n_{u} L}{T_{u}^{2}}$, where $L$ is the divertor connection length \cite{stangeby2000plasma}. This figure shows the closed geometry is able to access detachment and low heat loads at lower collisionalities, a significant benefit for core-edge integration. Again these differences seem to lessen at higher powers, with the 3 MW cases showing more than a factor of 4 upstream collisionality at the threshold of detachment, and the 12 MW cases showing less than a 10 \% difference. Figure \ref{fig:BaffleFluxEvolution} also shows the closed geometry tends to have slightly higher peak loads when attached for the same density than the open geometry. This is due to the enhanced particle flux at the target for the same degree of detachment, which enhances both sheath transmission and surface recombination loads. This further highlights the idea that the closed case is not uniformly better when it comes to accessing good target conditions.

\section{Power Balance}

\begin{figure}[ht]
\begin{subfigure}{.5\textwidth}
  \centering
  \includegraphics[width=\linewidth]{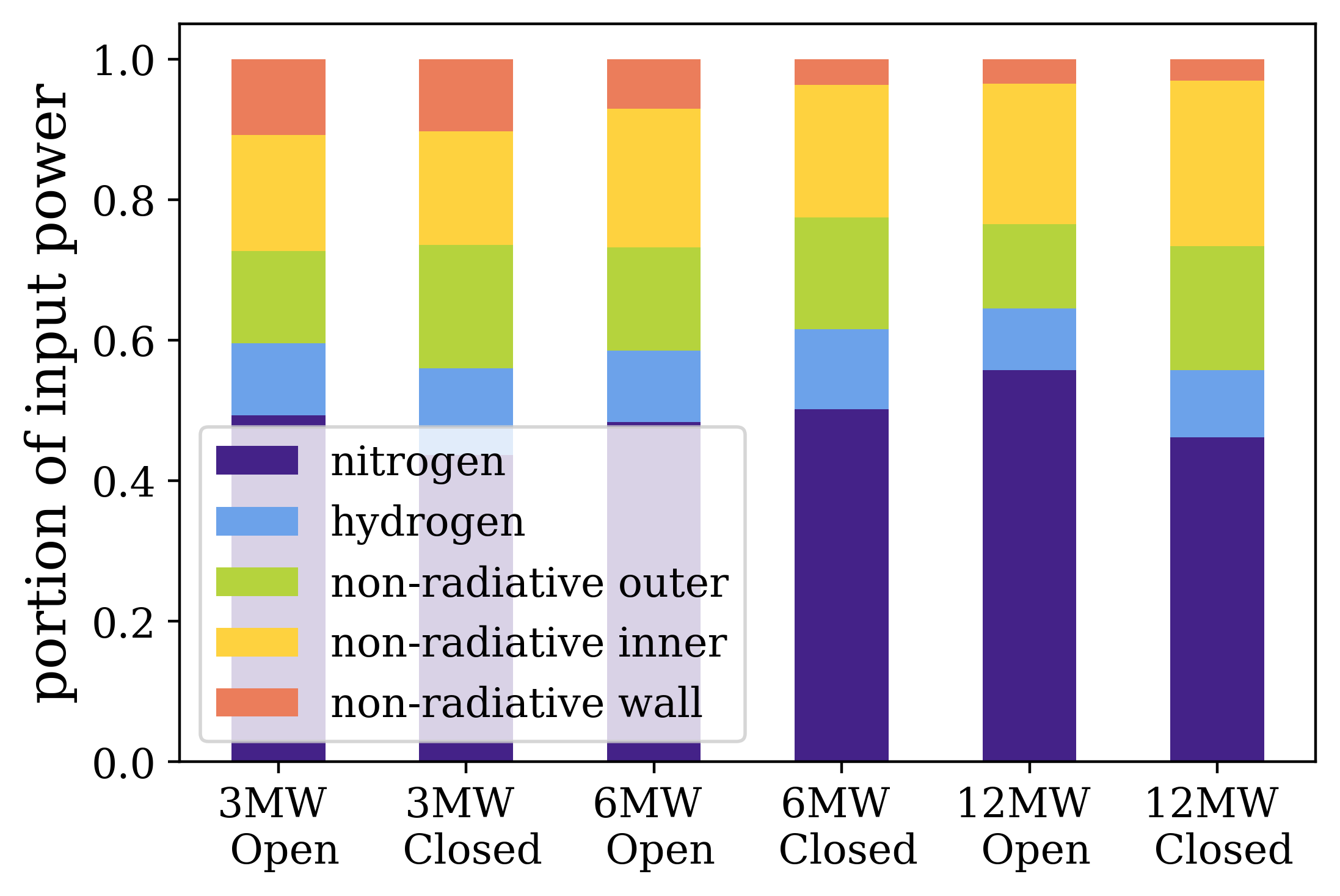}  
\caption{}
\label{fig:heat_source_type_baffle}
\end{subfigure}
\begin{subfigure}{.5\textwidth}
  \centering
  \includegraphics[width=\linewidth]{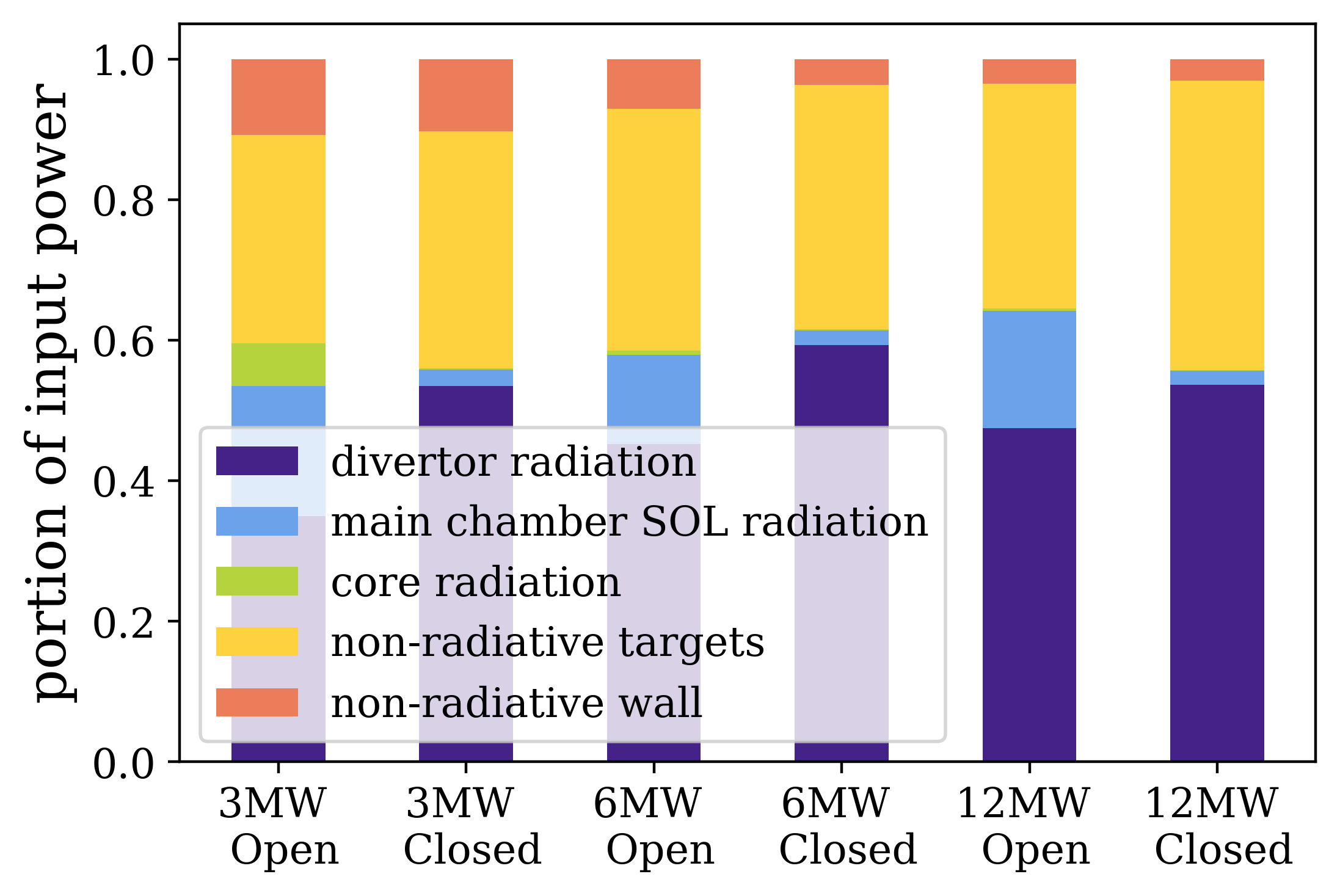} 
\caption{}
\label{fig:heat_location_baffle}
\end{subfigure}
\caption{The balance of total power for the threshold of
detachment simulations at three different input powers, for an open
and closed geometry. a) Shows the balance of power decomposed by sink type, and b) shows the power loss decomposed by region.}
\label{fig:powerBalance}
\end{figure}

To further understand the difference in detachment characteristics caused by baffling, it is useful to consider the power balance at the threshold of detachment for each geometry. Figure \ref{fig:heat_source_type_baffle} shows the sinks of total power decomposed by sink type, including the total direct cross-field heat flux deposited to the walls (not including radiation, labelled `non-radiative wall'), the total direct poloidal heat flux deposited on the inner and outer targets (not including radiation, labelled `non-radiative inner' and `non-radiative outer'), the power radiated by nitrogen, and the power radiated by hydrogen. From this figure, perhaps surprisingly one can see that the amount of power dissipated by the different loss channels remains roughly constant between geometries. 

However, the simulations at high powers show some noticeable differences. In particular, less particle-driven heat flux is directed radially towards the walls at higher powers. This is consistent with radial transport being reduced at higher powers for the same transport coefficients. Additionally, as power is increased, slightly more power is incident on the inner targets in relative to the outer targets. At 3 MW, the in/out power flux ratio for the open geometry is roughly 1:1, whereas at 12 MW, this ratio is roughly 2:1. It is important to note that this difference in in/out target loads is not due to a difference in power crossing the separatrix, but is instead due to a change in the ratio of dissipated power along the inner and outer legs.

In addition to decomposing power by sink type, the power balance has been decomposed by different regions. These regions include the core (inside the separatrix), main chamber SOL (outward from the separatrix, upstream of the plasma x-points), divertor (regions downstream of the x-points), and the power leaving the radial and poloidal boundaries towards the wall and targets, and the results are shown in Figure \ref{fig:heat_location_baffle}. From this figure, a clear difference between the geometries is that significantly more power is radiated in the main chamber in the open geometry, relative to the closed geometry. Moreover, this difference is consistent at all power levels; the open geometry radiates roughly 3 times more power in the divertor than the main chamber, whereas the closed geometry radiates roughly 25 times more power in the divertor than the main chamber.

\section{Nitrogen Radiation and the Killer Flux Tube}\label{sec:Nrad}

Noting the aforementioned differences in detachment access, one may wonder how the closed divertor is able to access detachment easier at lower powers, and why this benefit seems to wane at high powers. To study this difference, it is useful to consider the power balance along the killer flux tube, since detachment is characterised by a drastic reduction in the parallel heat flux density in this flux tube. And, since dissipation is dominated by nitrogen radiation, then any differences between the two geometries is due to a difference in efficiency of nitrogen radiation, or a difference in power entering the flux tube.

\begin{figure}[ht]
\begin{subfigure}{.32\textwidth}
  \centering
  \includegraphics[width=\linewidth]{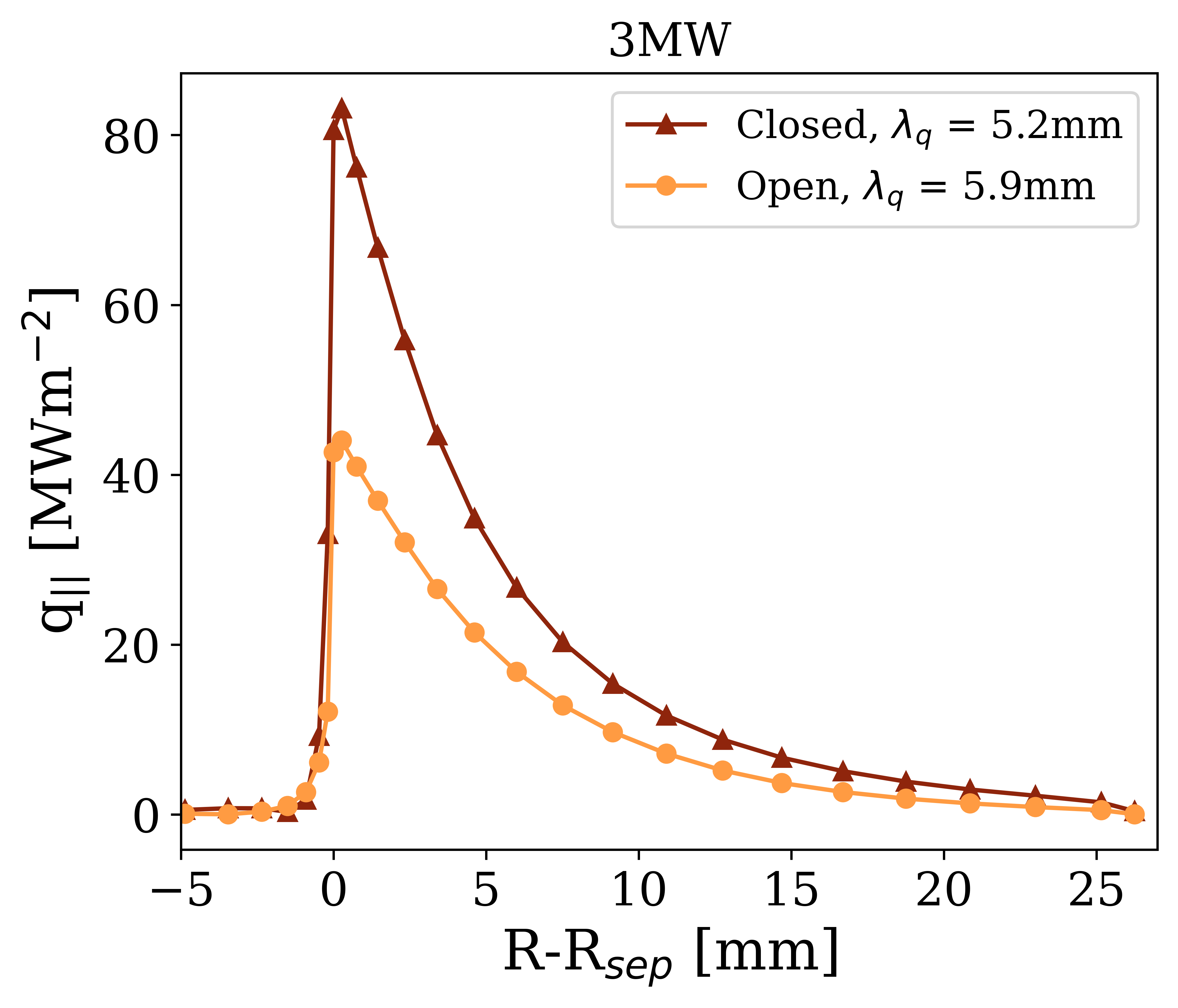}  
\caption{}
\label{fig:baffle_Heat_faloff3MW}
\end{subfigure}
\begin{subfigure}{.32\textwidth}
  \centering
  \includegraphics[width=\linewidth]{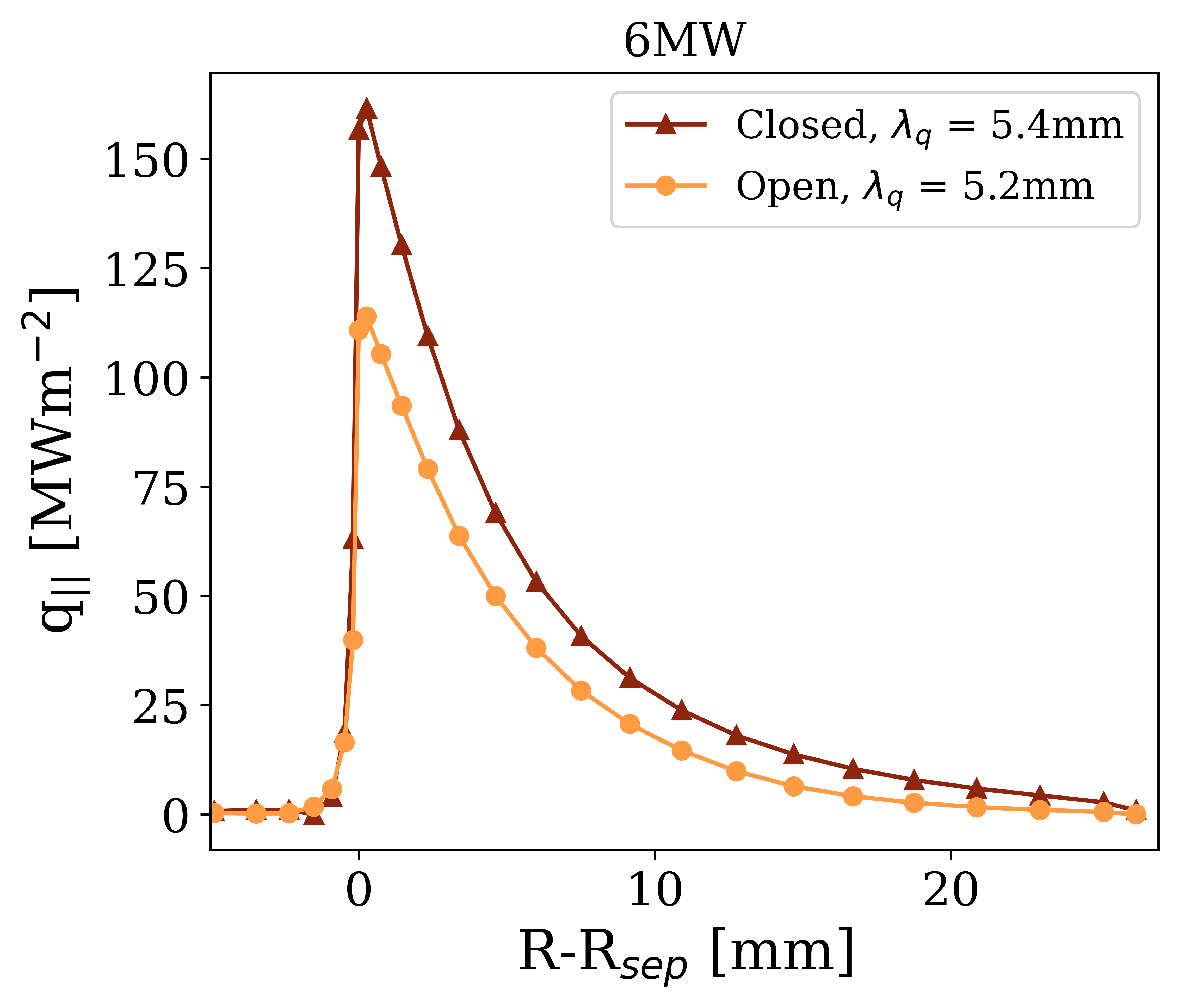} 
\caption{}
\label{fig:baffle_Heat_faloff6MW}
\end{subfigure}
\begin{subfigure}{.32\textwidth}
  \centering
  \includegraphics[width=\linewidth]{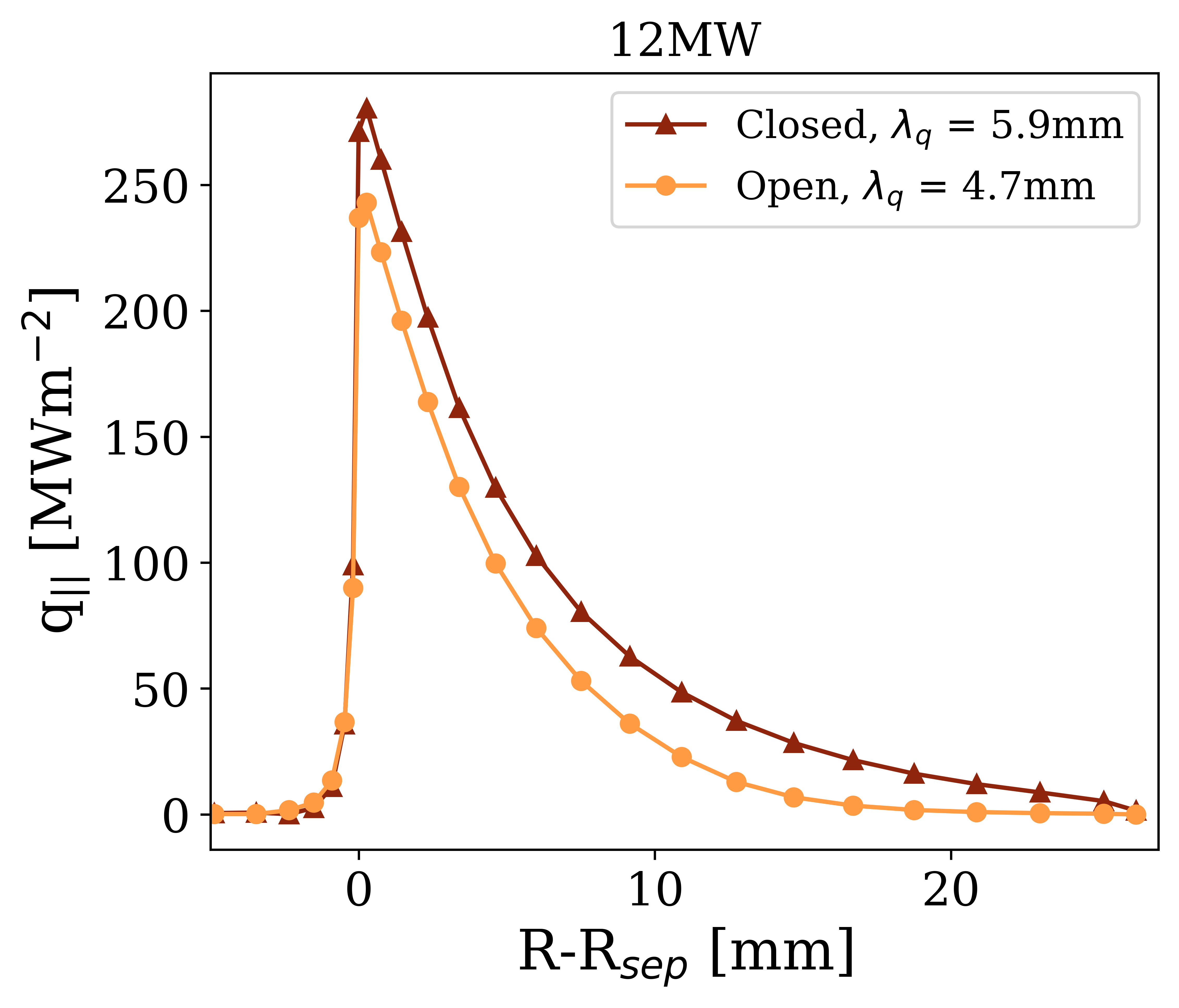} 
\caption{}
\label{fig:baffle_Heat_faloff12MW}
\end{subfigure}
\caption{The parallel heat flux density at the lower outer divertor entrance, plotted as a function of cross-field distance from the separatrix (mapped to the midplane). The data is for open and closed MAST-U geometries, at a) 3 MW, b) 6 MW, and c) 12 MW of input power.}
\label{fig:baffle_Heat_faloff}
\end{figure}

Comparing power balance along the killer flux tube at detachment, one of the most apparent differences caused by baffling is the heat flux entering this flux tube. Due to the presence of strong radiation in the main chamber SOL, the parallel heat flux density at the x-point of the killer flux tube is roughly 2 times lower in the open geometry than in the closed geometry at 3 MW. This can be seen in Figure \ref{fig:baffle_Heat_faloff}, which shows the parallel heat flux profiles at the x-point for both geometries at all three power levels. Interestingly, this difference in heat flux is reduced at high powers, due to a narrowing in heat flux width for the open case, a widening of the heat flux width in the closed divertor, and the main chamber radiation moving outwards to the far SOL in the open geometry, away from the killer flux tube.

This variation in heat flux, however, does not explain why the closed case accesses detachment easier, and in fact would be an argument for the opposite to be true. So what enables the closed divertor to dissipate more power at the same density? Surprisingly, the presence of other non-nitrogen sources and sinks is negligible, and nitrogen dissipates the same fraction of the power in the killer flux tube of each geometry.

Consequently, it seems the efficiency of nitrogen radiation is the main cause of the differences in detachment access. In particular, the closed case has three ways of boosting nitrogen radiation. First, due to the differences in heat fluxes, the closed geometry is able to maintain higher upstream temperatures for the same target conditions. This higher temperature means a higher pressure for a given density, and according to models such as the DLS model \cite{lipschultz2016sensitivity,Cowley_2022}, it is the upstream pressure (not density) that determines the strength of impurity radiation. This is because, if pressure is conserved above the cold end of the radiating region, then the upstream pressure entirely determines the density at a particular temperature, such as 5 eV. Hence, the upstream pressure determines the density in the effective radiation region of an impurity, which for the impurity used is around 2-20 eV. In fact, not only does the open geometry show inherently lower upstream temperatures, this geometry also has significantly more static electron pressure loss along the killer flux tube, which reduces the radiative power of nitrogen.

Finally, nitrogen radiation can be impacted by geometry through the volume available for radiation. In fact, from the previous power balance analysis, it was shown that power losses in the closed geometry are more confined to the divertor region, near the targets. Since the equilibrium is a Super-X, the total magnetic field is lower in this region near the targets, and the magnetic flux tubes are expanded and have more effective volume available for radiation. In fact, the radiation-averaged magnetic field strength is 0.36 T in the closed geometry, and 0.48 T in the open geometry at the threshold of detachment at 3 MW input power. Essentially, the closed geometry is leveraging the Super-X more effectively by having more localised radiation near the targets. 

This, coupled with the differences in temperature and pressure, is why the closed divertor is able to radiate more with the same upstream density. Moreover, some of these factors tend to reduce at high powers; the upstream temperatures of the two grids are more similar, the open case exhibits less pressure loss at higher powers. This explains why less of a difference in detachment threshold is seen at these powers. 

\section{Hydrogen Transport and Fuelling}

When discussing the prospective benefits and disadvantages of a closed divertor, baffling is thought to have drastic impacts on hydrogen transport. In particular, one would expect trapping of neutrals to differ significantly. In the open geometry, the expectation is that neutrals recycled from the targets may freely leave the divertor region, and so the proportion of recycled neutrals which are ionised in the divertor should be low. In contrast, one may expect the neutral trapping (the proportion of recycled neutrals ionised in the divertor) to be much higher in a closed geometry, since neutrals may not escape the divertor as easily. To quantify this we may calculate the neutral trapping:

\begin{equation}
    \eta = \frac{S_{ion,div}}{\Gamma_{div}},
\end{equation}

Where $\eta$ is the neutral trapping of a divertor, $S_{ion,div}$ is the total ionisation source in the divertor region, and $\Gamma_{div}$ is the total recycled particle flux from said region's divertor targets. Indeed, at all powers the trapping of neutrals is roughly a factor of 2 higher in the closed case than the open case at the threshold of detachment for the lower outer divertor.

\begin{figure}
\begin{subfigure}{.5\textwidth}
  \centering
  \includegraphics[width=\linewidth]{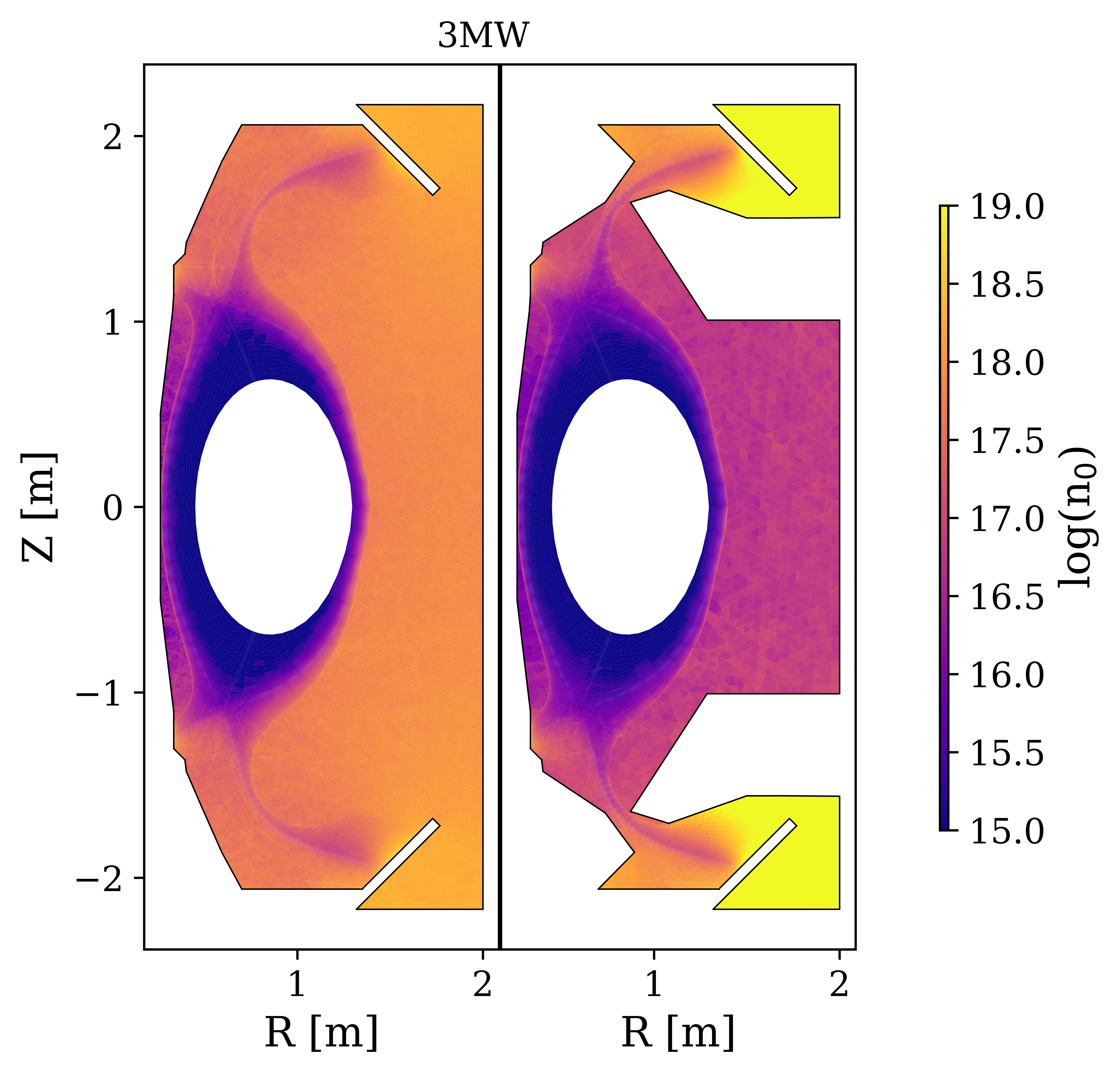}  
\caption{}
\label{fig:Ddensity_Baffle}
\end{subfigure}
\begin{subfigure}{.5\textwidth}
  \centering
  \includegraphics[width=\linewidth]{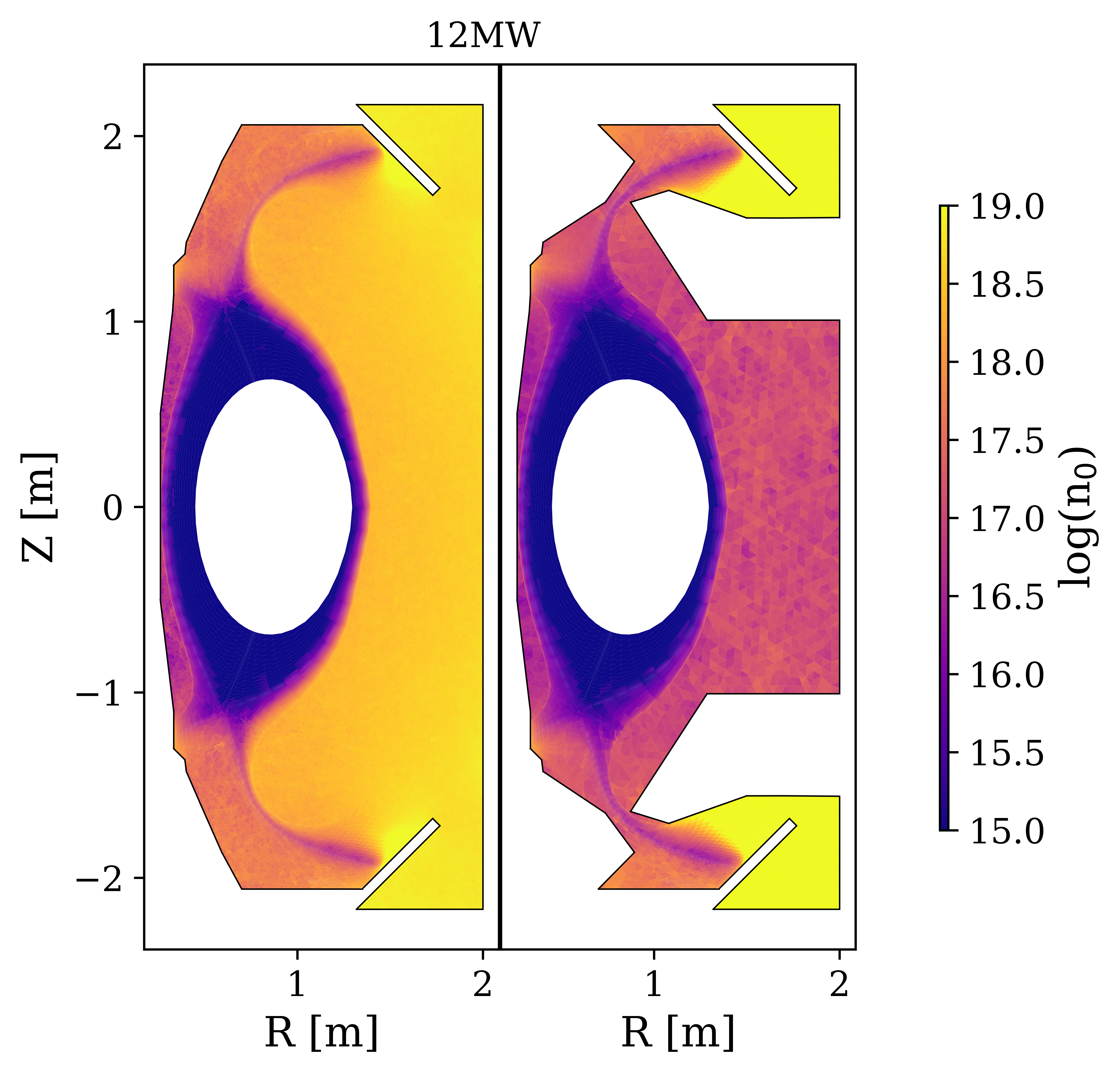} 
\caption{}
\label{fig:Ddensity_Baffle12MW}
\end{subfigure}
\caption{The 2D profiles of total neutral deuterium density for closed and open baffled MAST-U Super-X geometries, with a) 3 MW and b) 12 MW of input power.}
\label{fig:Ddensity_Baffle_All}
\end{figure}

This difference in neutral trapping leads to a significant difference in neutral density distributions and neutral compression between the two geometries. Figures \ref{fig:Ddensity_Baffle} and \ref{fig:Ddensity_Baffle12MW} show the 2D neutral density profiles for 3 MW and 12 MW input power respectively. In these figures the open baffled geometry has a relatively uniform neutral density profile, as deuterium atoms and molecules are free to move around most of the chamber. As a result, the compression, or neutral density ratio in the divertor compared to the midplane is only 2 at 12 MW. In contrast, the closed geometry shows a compression of nearly 450 at 12 MW.

When comparing low and high powers, one may assume that an open geometry would show significantly more neutral trapping at high powers. This is because the plasma may itself act as a baffle, since its high temperatures would quickly ionize any recycled neutrals and redirect them towards the targets. Inspecting Figure \ref{fig:Ddensity_Baffle12MW}, we see the angle of the targets means the recycled neutrals are not directed back into the plasma, but can instead escape around the low-field side of the divertor leg and fill the low-field side chamber. However, the plasma does clearly act more like a baffle at 12MW, and there is a significant difference in neutral density on the high and low-field sections of the chamber due to reduced neutral transport across the plasma. This reduced communication of neutrals between the inner and outer chambers likely contributes to the increased inner target heat flux at higher powers in the open geometry.

An important and obvious impact of this difference in neutral transport is pumping. In the closed case, for a given upstream density there is a much stronger buildup of neutral pressure in the divertor region, and as a result pumping neutrals out of the divertor is significantly easier. In fact, at 3 MW the pumping rate is enhanced by a factor 7 over the open case. At 12MW, this enhancement of neutral pumping increases to a factor 10 going from the open to closed geometry. 

\begin{figure}[ht]
\begin{subfigure}{.32\textwidth}
  \centering
  \includegraphics[width=\linewidth]{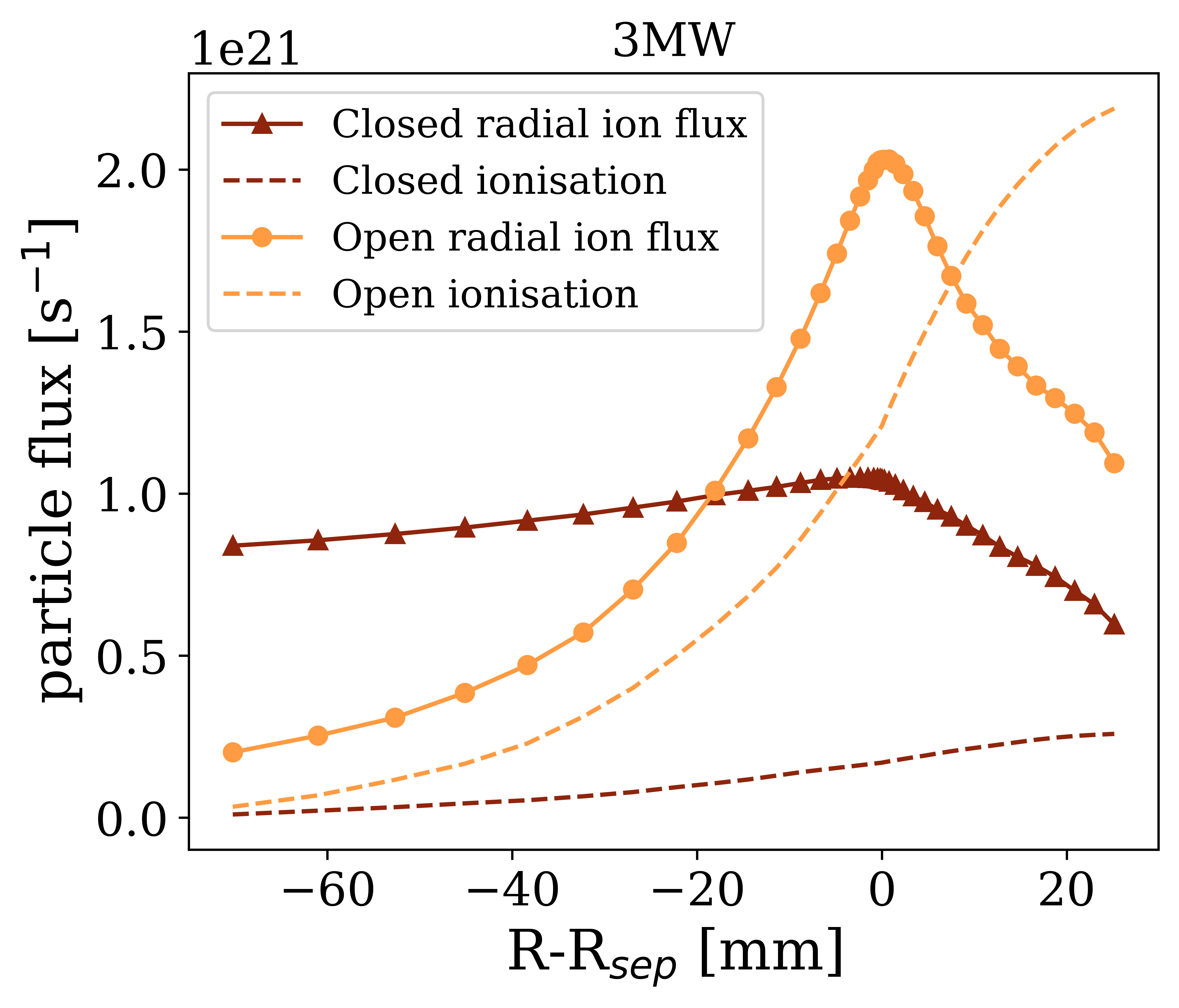}  
\caption{}
\label{fig:upstream_particle_flux3MW}
\end{subfigure}
\begin{subfigure}{.32\textwidth}
  \centering
  \includegraphics[width=\linewidth]{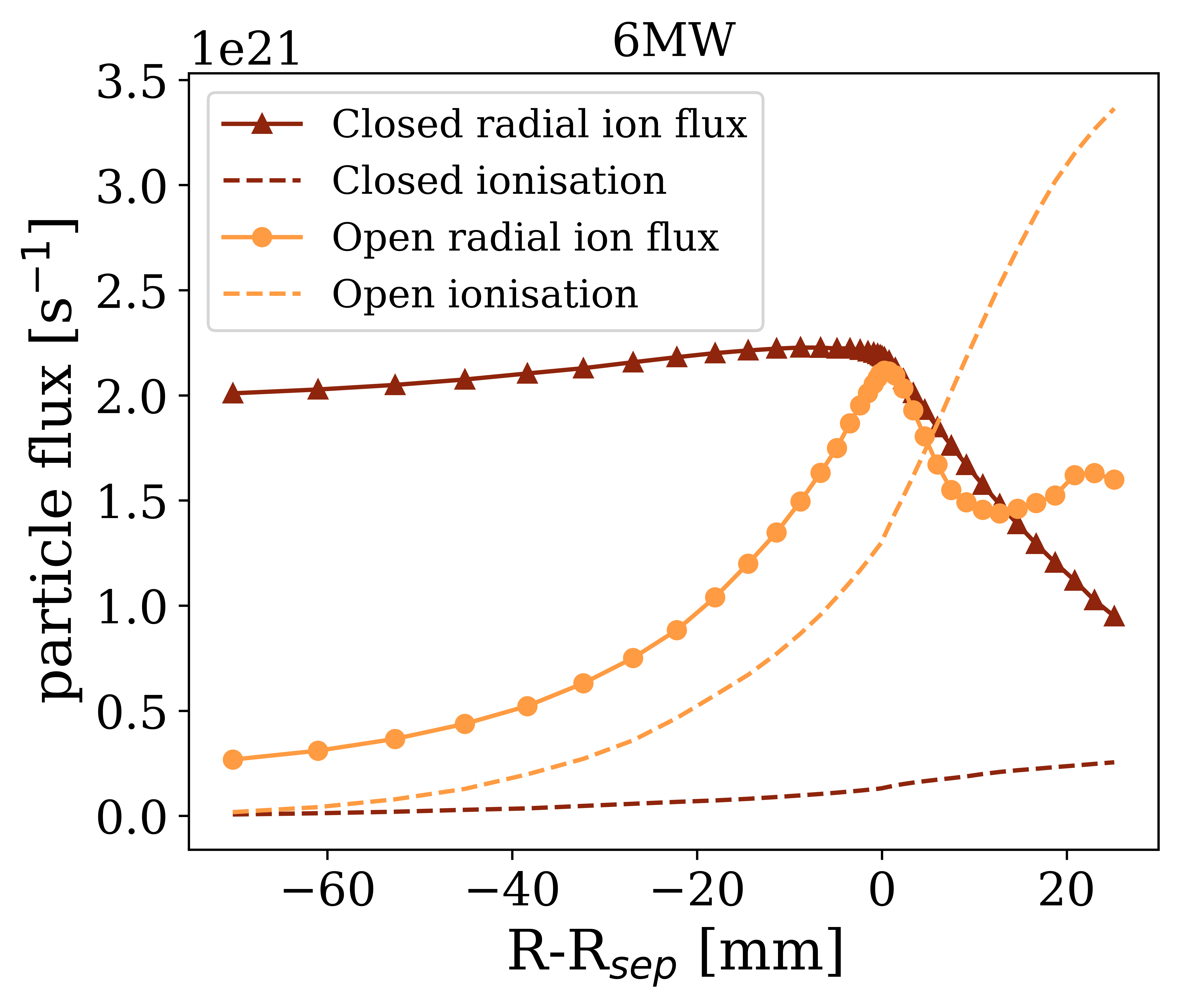} 
\caption{}
\label{fig:upstream_particle_flux6MW}
\end{subfigure}
\begin{subfigure}{.32\textwidth}
  \centering
  \includegraphics[width=\linewidth]{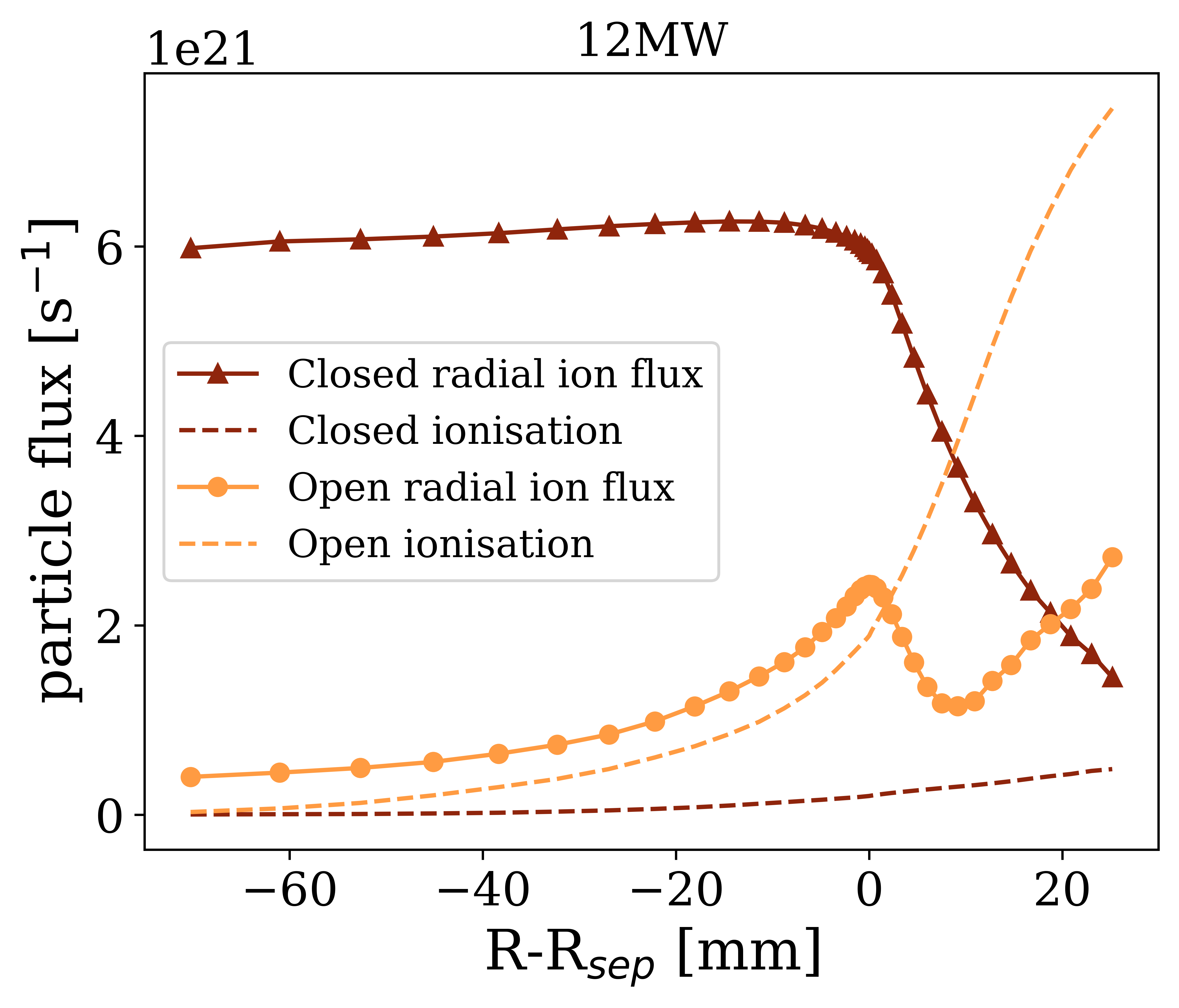} 
\caption{}
\label{fig:upstream_particle_flux12MW}
\end{subfigure}
\caption{The radial ion flux through the outer midplane grid cells, plotted as a function of cross-field distance from the separatrix, for open and closed MAST-U Super-X geometries at the threshold of detachment, at a) 3 MW, b) 6 MW, and c) 12 MW input power. Also shown is the cumulative summed ionisation particle source in these grid cells. }
\label{fig:baffle_flux}
\end{figure}

\begin{figure}
\begin{subfigure}{.32\textwidth}
  \centering
  \includegraphics[width=\linewidth]{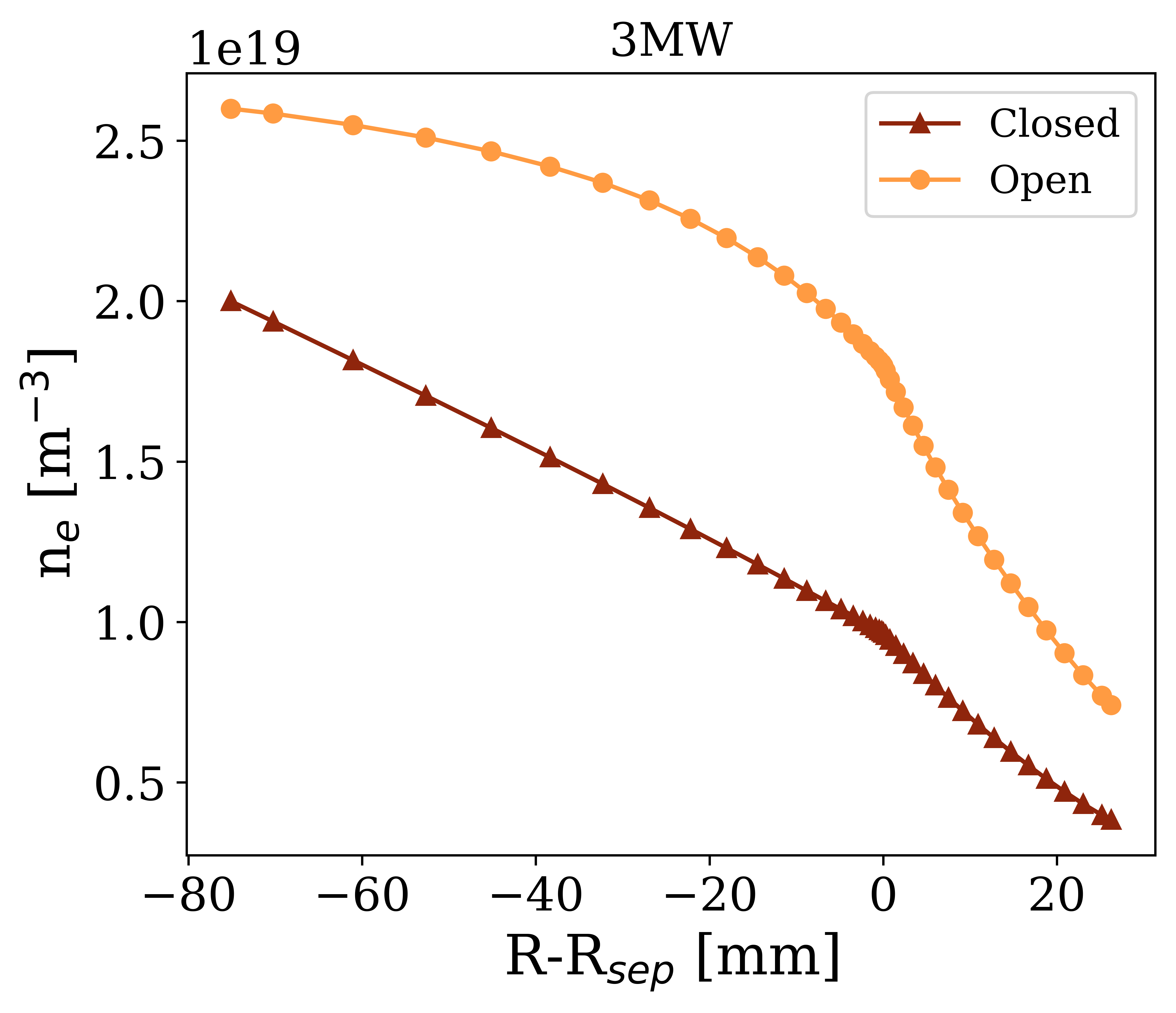}  
\caption{}
\label{fig:baffle_densUpstream3MW}
\end{subfigure}
\begin{subfigure}{.32\textwidth}
  \centering
  \includegraphics[width=\linewidth]{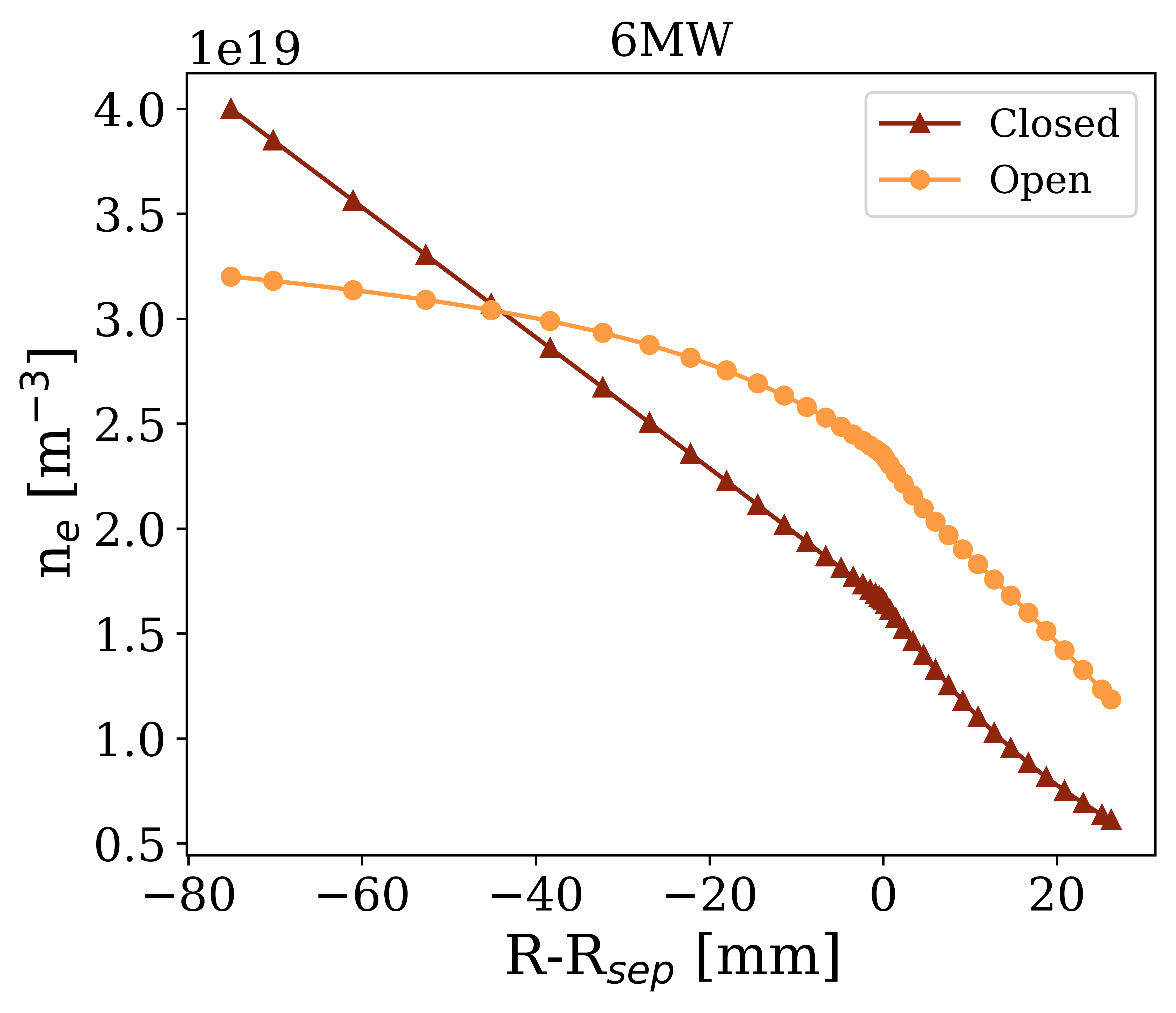} 
\caption{}
\label{fig:baffle_densUpstream6MW}
\end{subfigure}
\begin{subfigure}{.32\textwidth}
  \centering
  \includegraphics[width=\linewidth]{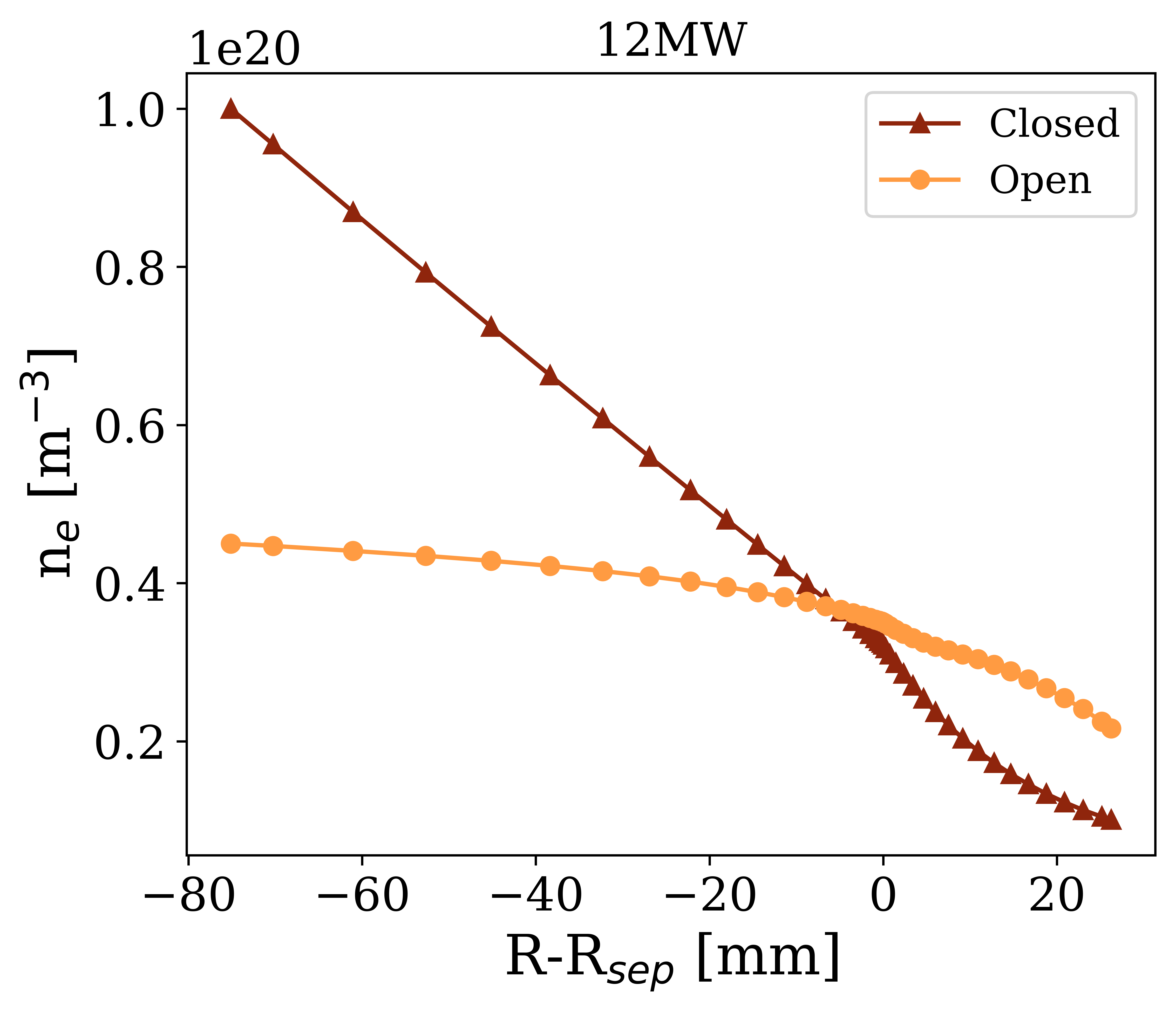} 
\caption{}
\label{fig:baffle_densUpstream12MW}
\end{subfigure}
\caption{The outer midplane density, plotted as a function of cross-field distance from the separatrix, for open and closed MAST-U Super-X geometries at the threshold of detachment, at a) 3 MW, b) 6 MW, and c) 12 MW input power.}
\label{fig:baffle_densUpstream}
\end{figure}

Through this extreme difference in neutral trapping and pumping, the presence of baffling strongly impacts particle fuelling and midplane profiles. Specifically, the open case, in which recycled neutrals are able to fill the main chamber volume, the plasma SOL is fuelled much more through recycling compared to the closed case. As a result, for similar densities, the open geometry requires a significantly lower particle source from the core to achieve a given separatrix density. This can be seen in Figure \ref{fig:baffle_flux}, which shows the radial particle flux and cumulative ionisation particle source at the midplane for all geometries and powers at the threshold of detachment. These F
figures show a significant ionisation source in the SOL in the open case, whereas the closed case is entirely fuelled by particle injection through the core. The lower input particle flux in the open case leads to a much flatter density profile, which becomes even flatter at higher powers, as can be seen in Figures \ref{fig:12MW_ne_Thresh} and \ref{fig:baffle_densUpstream}.

\begin{figure}[ht]
\begin{subfigure}{.32\textwidth}
  \centering
  \includegraphics[width=\linewidth]{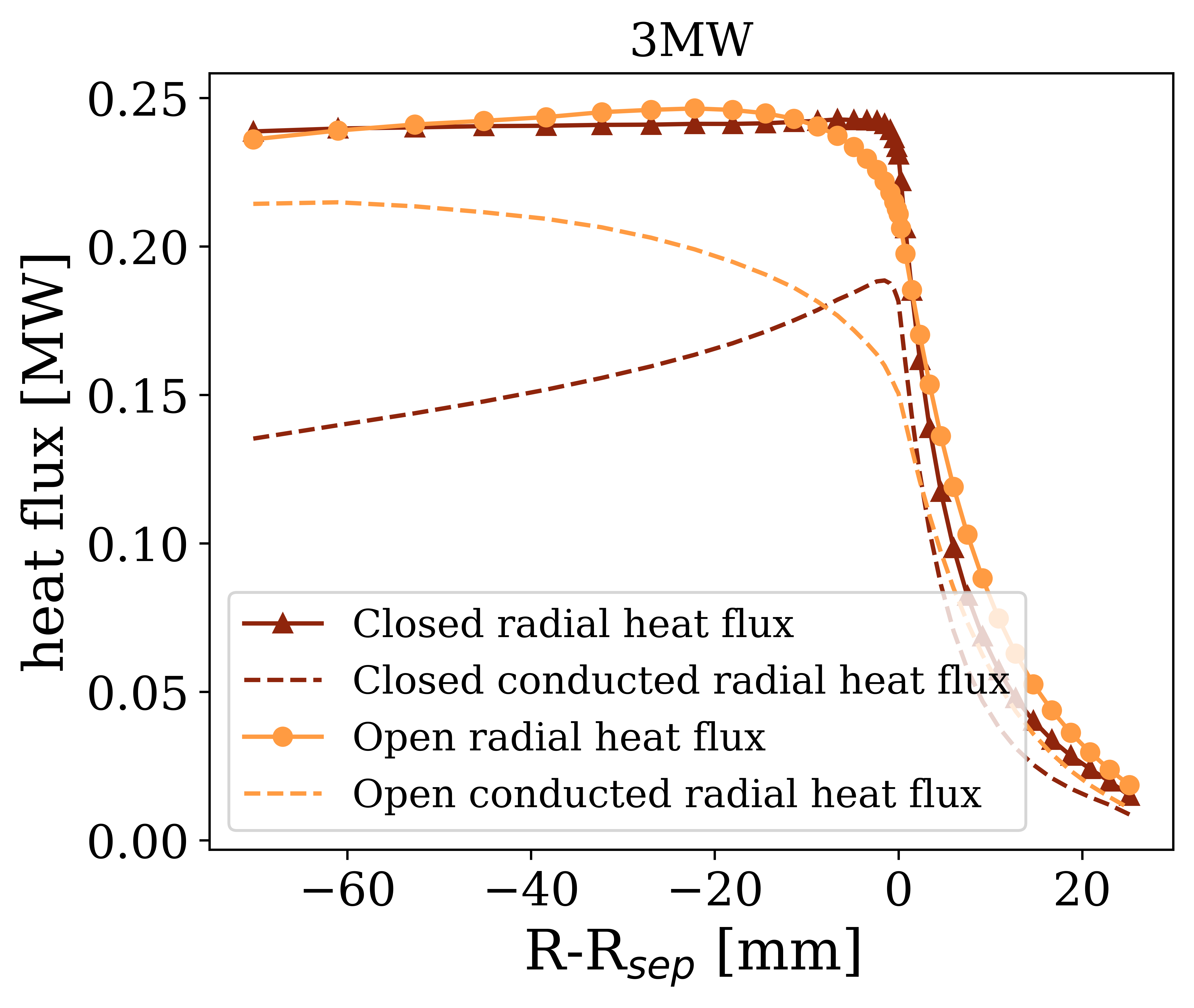}  
\caption{}
\label{fig:radHeatflux3MW}
\end{subfigure}
\begin{subfigure}{.32\textwidth}
  \centering
  \includegraphics[width=\linewidth]{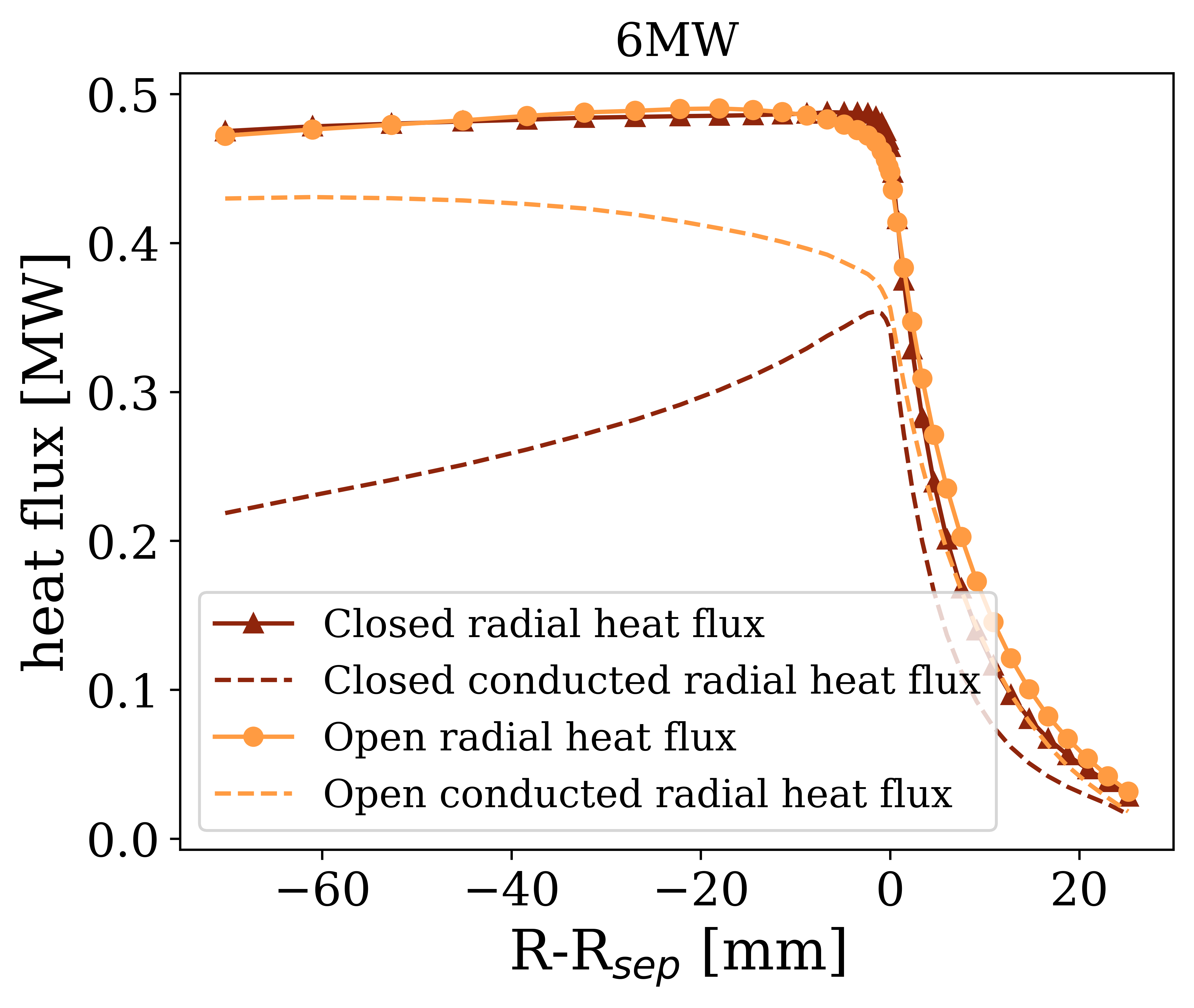} 
\caption{}
\label{fig:radHeatflux6MW}
\end{subfigure}
\begin{subfigure}{.32\textwidth}
  \centering
  \includegraphics[width=\linewidth]{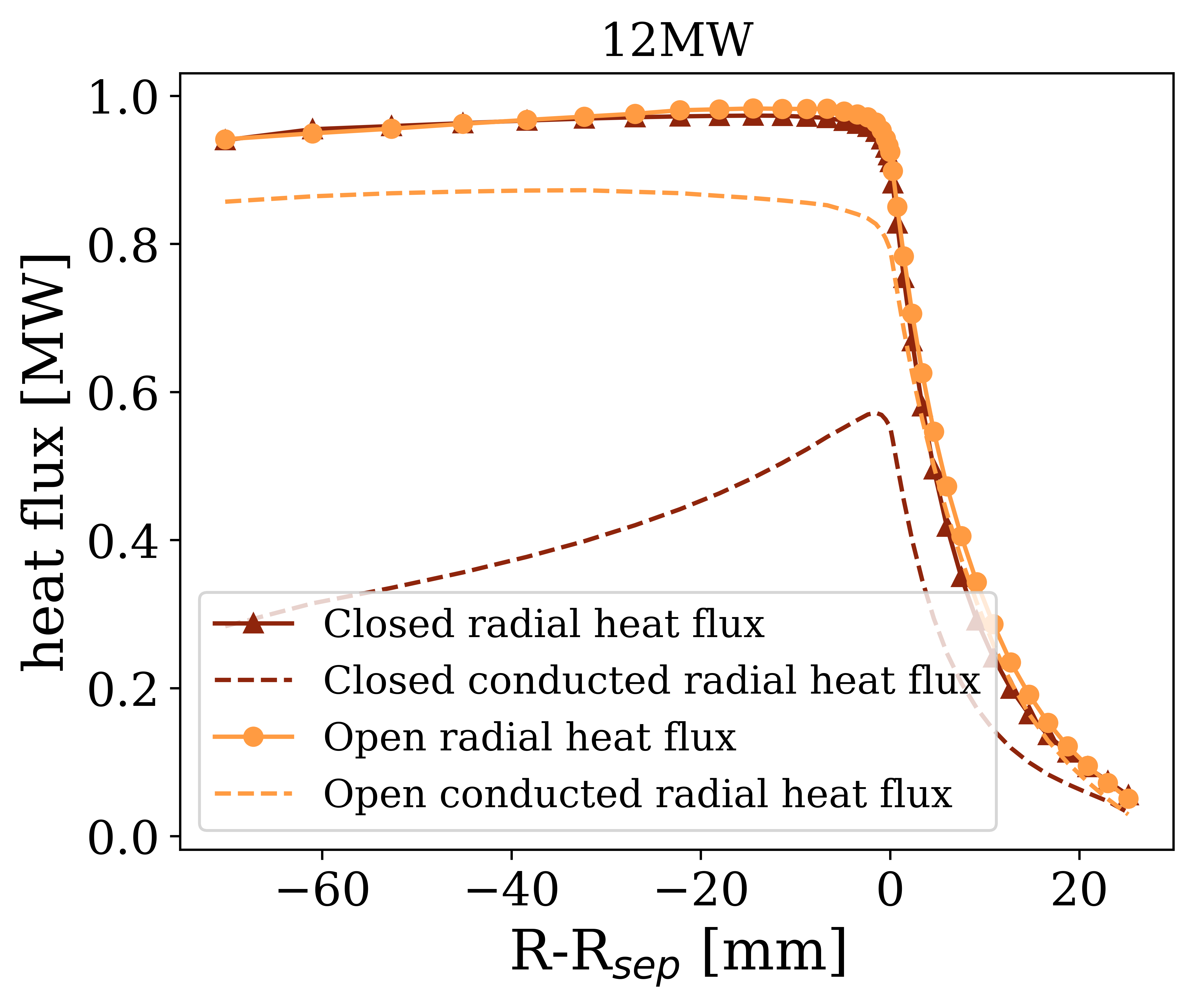} 
\caption{}
\label{fig:radHeatflux12MW}
\end{subfigure}
\caption{The total radial heat flux and conducted radial heat flux through the grid cells at the outer midplane, plotted as a function of cross-field distance from the separatrix, for open and closed MAST-U Super-X geometries at the threshold of detachment, at a) 3 MW, b) 6 MW, and c) 12 MW input power.}
\label{fig:radHeatflux}
\end{figure}

Finally, the differences in fuelling between the open and closed geometries also contribute to variations in radial profiles of heat flux. Figure \ref{fig:radHeatflux} shows the outer midplane radial heat flux profiles for all geometries and power levels at the threshold of detachment. Also shown in this figure is the conducted portion of the radial heat flux. From this figure, it is clear to see that while the total radial heat flux profiles are similar between the two grids, the amount of this heat flux in the conduction channel is different. In particular, the closed geometry tends to have less input power in the conductive channel, and more in the convective due to the significantly higher input particle fluxes from pellet fuelling. Moreover, at higher powers the heat flux crossing the separatrix is significantly convected in the closed case, and at 12 MW the separatrix conducted radial heat flux is $\approx$ 50\% lower in the closed case compared with the open geometry. This may explain why - at high powers - the heat flux width of the closed divertor is larger than that of the open.

\section{Parallel Flows}

\begin{figure}[ht]
\begin{subfigure}{.49\textwidth}
  \centering
  \includegraphics[width=\linewidth]{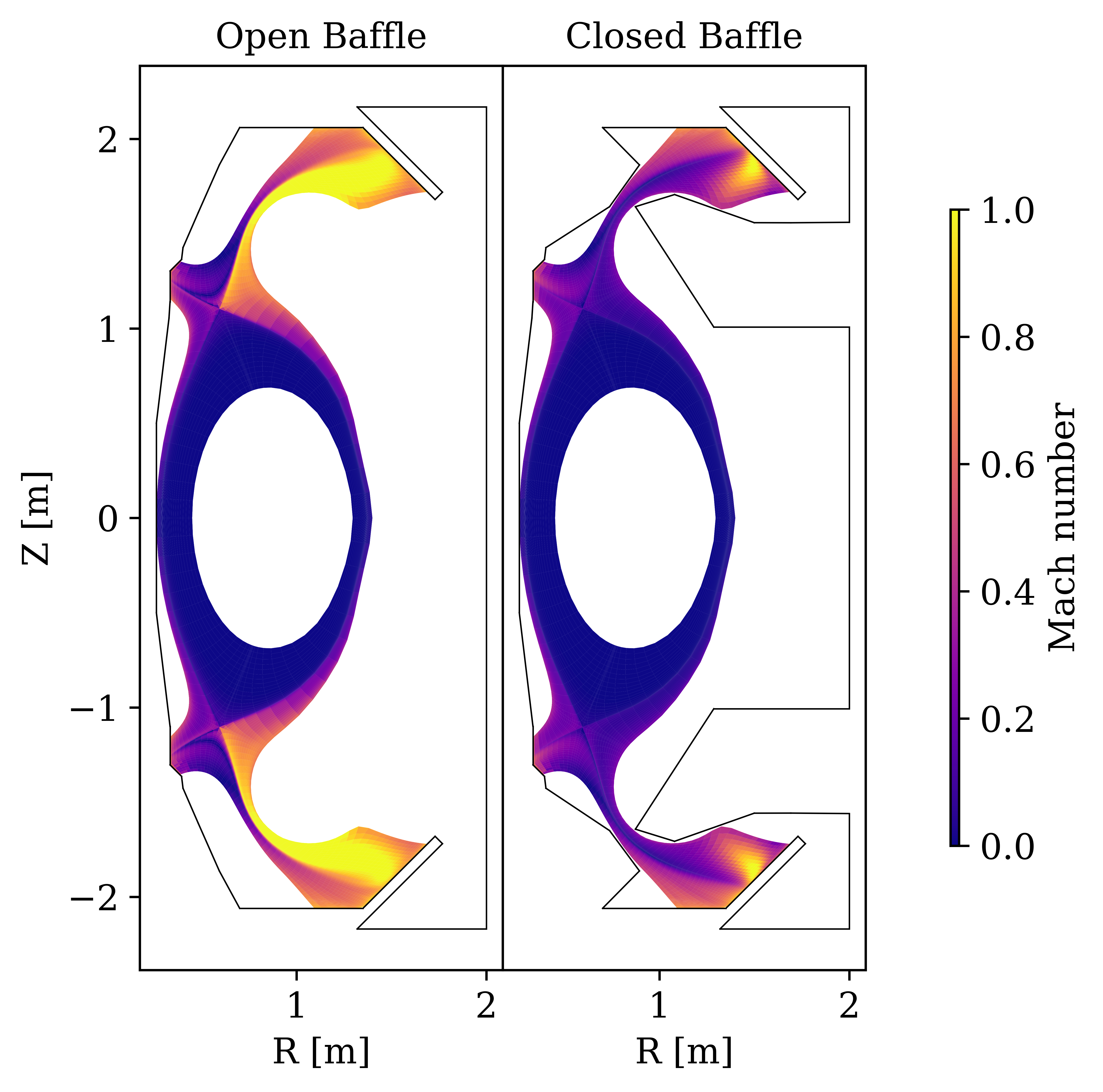}  
\caption{}
\label{fig:3mwmach}
\end{subfigure}
\begin{subfigure}{.49\textwidth}
  \centering
  \includegraphics[width=\linewidth]{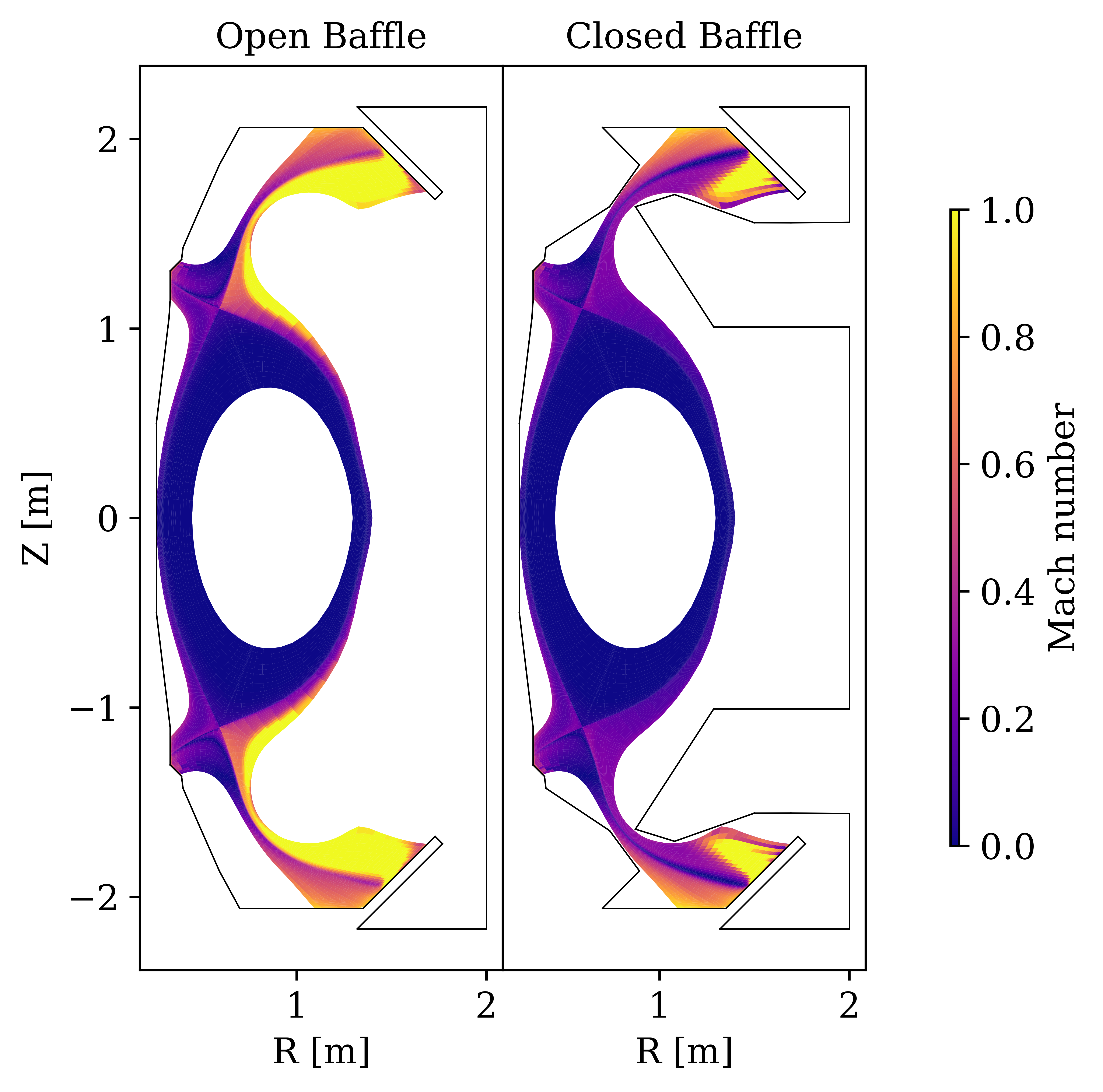} 
\caption{}
\label{fig:12mwmach}
\end{subfigure}

\caption{The deuterium mach number profiles for a) 3 MW and b) 12 MW simulations at the threshold of detachment; for open and closed MAST-U Super-X geometries.}
\label{fig:mach}
\end{figure}

In previous studies, a key difference seen between open and closed divertors has been the presence of strong flows in the main chamber. To confirm this, the mach number for deuterium ions is shown in Figure \ref{fig:mach}, for 3 MW (a) and 12 MW (b) simulations at the threshold of detachment. For each power, the simulated profiles for the open geometry is shown on the left, and the closed geometry is shown on the right.

Figure \ref{fig:mach} indeed confirms the presence of much stronger flows in the main chamber in the open geometry. This is caused by the strong ionisation upstream causing sources of ion flux. Unsurprisingly, these strong flows also contribute to the reduction in electron conducted heat flux at the divertor entrance, since a higher fraction of the heat flux is convected.

Because of the high mach numbers upstream, much of the pressure is dynamic in the open geometry. Since this ram pressure is strongly impacted by the geometry of flux expansion, the total flux expansion of the Super-X reduces the total pressure in the killer flux tube by nearly 50\% in the open geometry at 3 MW. For the closed geometry, with weaker upstream flows, flux expansion only reduces the total pressure by 10 \%. Additionally, in the open geometry the dynamic pressure increases significantly from midplane to the divertor chamber. This increase in dynamic pressure, and reduction in total pressure explains why the static electron pressure in the radiating region is much lower in the open divertor compared with the closed divertor. In Section \ref{sec:Nrad} it was found that pressure variation is a likely reason why the radiative power and detachment access of the two geometries is different. Hence, it seems the presence of strong parallel flows is crucial to understanding the aforementioned differences in detachment access. 

\section{Detachment Evolution}\label{sec:evolution}

\begin{figure}[ht]
  \centering
  \includegraphics[width=0.8\linewidth]{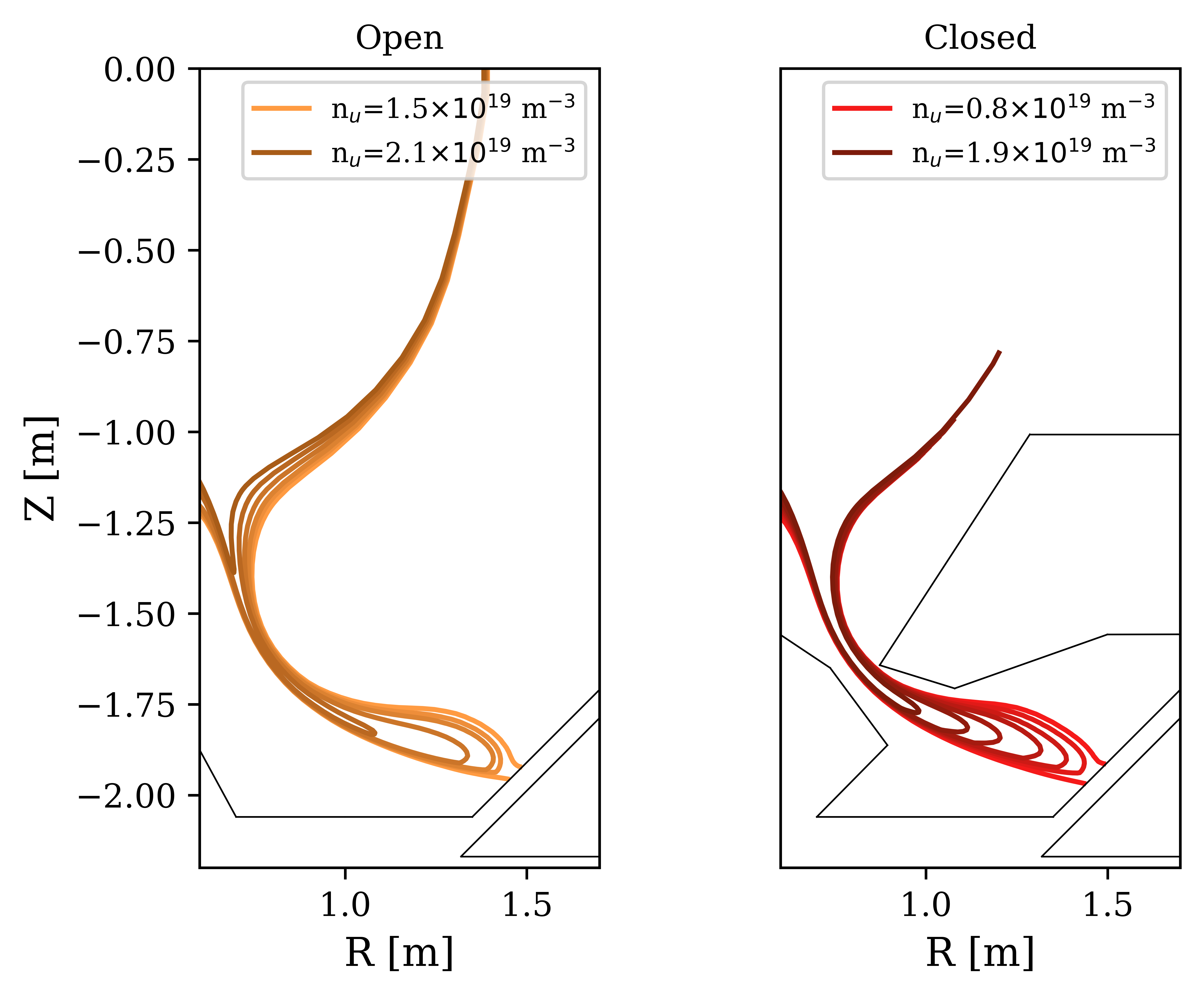} 

\caption{Contours of the 10 eV electron temperature, for simulations in the density scan in an open (left) and closed (right) MAST-U divertor at 3 MW input power. The density scans in the open scenario corresponds to a change in density of 40\%, and the closed scan corresponds to a change in density of 140\%. Darker shades indicate higher density simulations.}
\label{fig:contour}
\end{figure}

Finally, an important aspect of divertor performance is how the entire plasma profile evolves as it is pushed into an increasingly detached state. To investigate the impact of baffling on the evolution of the plasma solution, the 10 eV electron temperature contours have been determined for each simulation in the 3 MW density scan for the closed and open geometries. This contour has been chosen as 5-10 eV is often indicative of the transition from a high temperature plasma to the cold, low pressure detached region. These contours are shown in Figure \ref{fig:contour}, for simulations ranging from the last attached simulation until the simulations failed in deep detachment.

\begin{figure}[ht]
\begin{subfigure}{.49\textwidth}
  \centering
  \includegraphics[width=\linewidth]{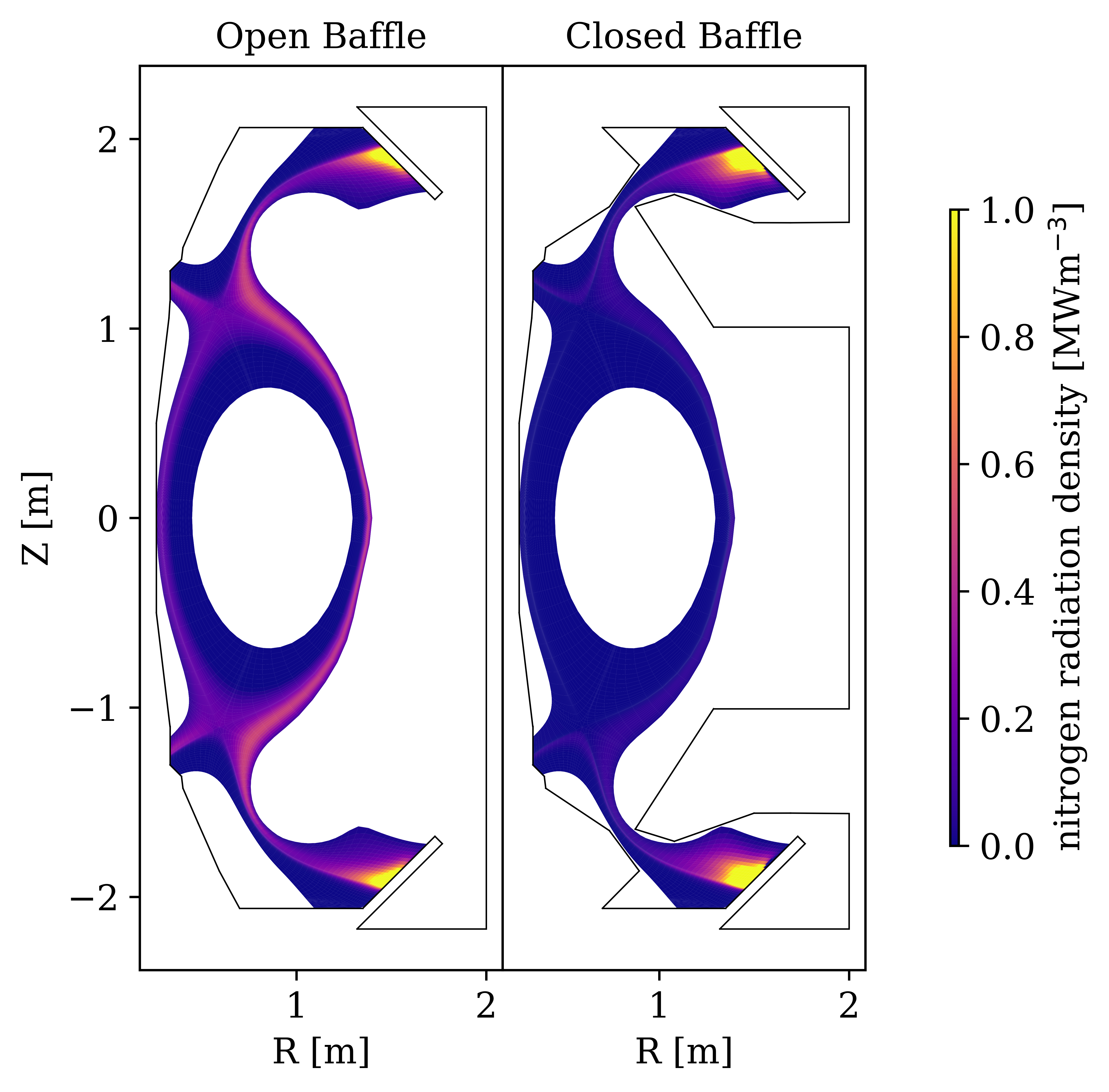}  
\caption{}
\label{fig:3MW_powerSink_Thresh_}
\end{subfigure}
\begin{subfigure}{.49\textwidth}
  \centering
  \includegraphics[width=\linewidth]{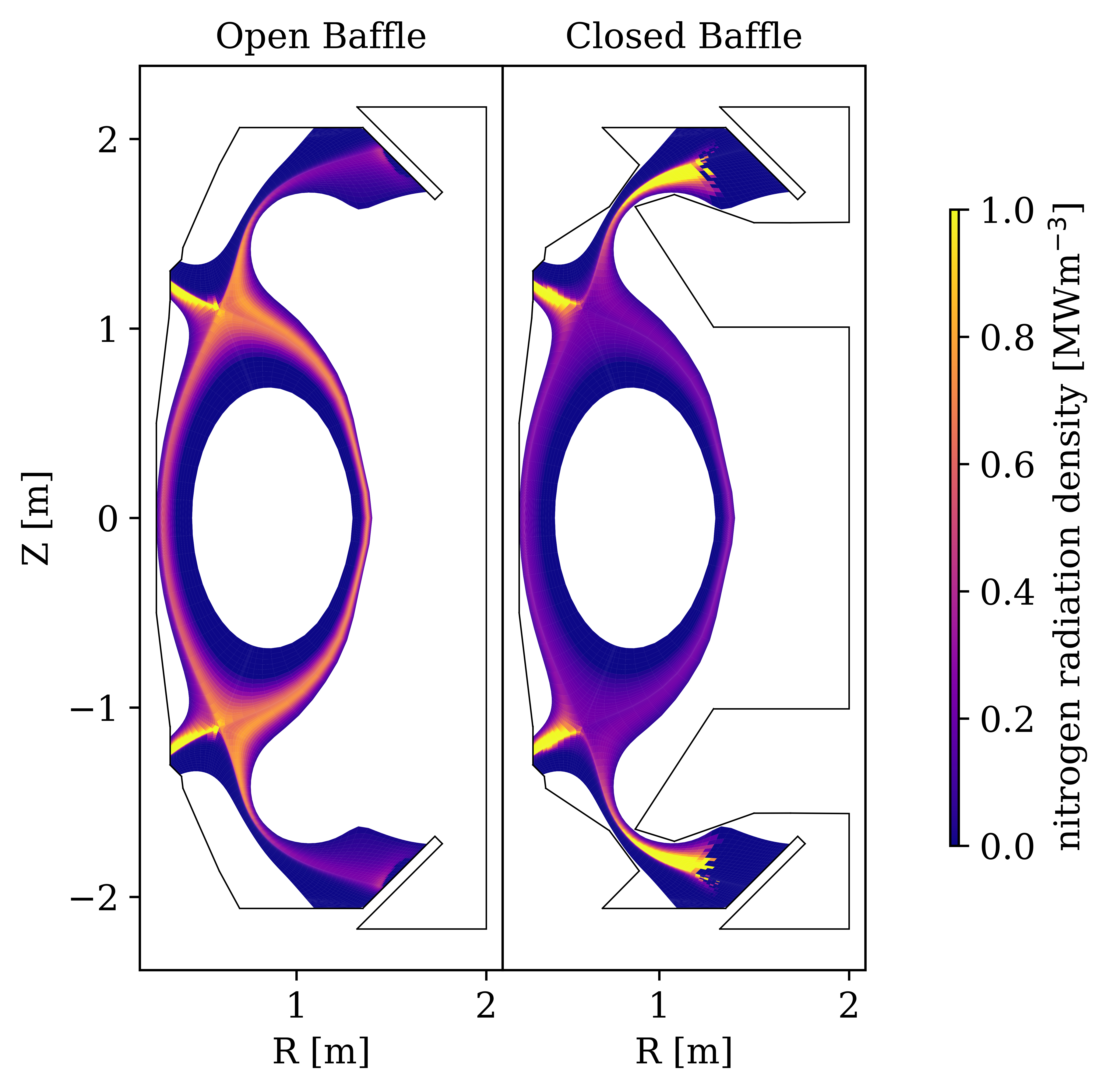} 
\caption{}
\label{fig:3MW_powerSink_Deep_}
\end{subfigure}

\caption{The nitrogen radiation profile for 3 MW closed and open baffled geometries, at the a) threshold of detachment and b) in deep detachment.}
\label{fig:powersinkEvolution}
\end{figure}

Figure \ref{fig:contour} shows several interesting features. As expected, the 10 eV contours for both geometries move further upstream as the density is ramped and an increasingly detached state is accessed. In the closed case, however, the temperature front only moves away from the target, and does not change significantly in width. In contrast, the temperature front in the open geometry moves both away from the target, and further inward to the core. In fact, in the most detached cases the 10 eV contour lies half way into the modelled SOL region at the x-point. This figure builds upon the initial conclusion in Section \ref{sec:DetachAcess} that the closed divertor has a larger window of detachment relative to the open divertor. In fact, these temperature contours show the open divertor has a much larger impact on the 2D plasma temperature profile with just a 40\% increase in density than the closed geometry, in which the density is increased by more than 140\%. 

This difference in temperature front evolution is most likely because the radiation in the open case is not in a localised volume near the targets. As discussed previously, the radiation in the open geometry occurs along the SOL and moves further inward and upstream as the divertor becomes more detached. This can be seen in Figure \ref{fig:powersinkEvolution}, which shows the nitrogen radiation profiles for the two geometries at the threshold of detachment, and in deeply detached states. Even in the threshold simulations, the radiation in the open geometry is more of a `mantle', whereas the closed geometry shows a highly localised radiation front near the target. As the simulations are pushed into deep detachment, the radiation in the open geometry mostly occurs in the main chamber, and is pushed further inwards, causing significant radiation around the x-point and inside the separatrix.  

In general, the evolution of both the temperature and radiation profiles show a stark contrast between the open and closed divertors. The closed geometry tends to show much more localised detachment fronts, which have an enhanced detachment window, and have less of an impact the upstream. It is important to emphasise is that the `radiating mantle' style of detachment in the open geometry is noted in literature to be generally less stable, and more prone to impacting confinement \cite{lipschultz1984marfe,lipschultz1987review,MartinGreenwald_2002}. The data shown here tends to be in agreement with this, and after increasing the density of the open simulations even further than those shown here, the simulations undergo complete radiative collapse.


\section{Discussion}

The SOLPS-ITER simulations in this work highlight several differences between closed and open divertor configurations. However, to truly understand what these differences mean for designing divertors, this work must be placed into context. Specifically, it is important to highlight what conclusions are likely generalisable or not, and which outcomes are beneficial or disadvantageous. 

In terms of generalised outcomes, this work consistently finds that strong baffling leads to several orders of magnitude heightened neutral compression in the divertor, and allows for a significantly more peaked density profile using core fuelling. These strong conclusions around hydrogen transport hold true at all power levels, and are consistently found in other experiments and simulations \cite{reimerdes2021initial}. There are small caveats with these conclusions, such as the fact that pellet fuelling does not perfectly penetrate into the core \cite{garzotti2011simulations}. What's more, the effective transport coefficients may change in reality between the two geometries, and the recycling of neutrals from the main chamber walls is not captured fully by SOLPS-ITER. Notwithstanding, one would still expect an open divertor to allow more neutrals to enter the main chamber, and lead to strong midplane fuelling through recycled neutrals and flatter density profiles.

The peaked density profile seen in closed simulations is likely a benefit for core-edge integration. This is because there is evidence to suggest that the operational Greenwald density is a limitation on separatrix density, not core density \cite{MartinGreenwald_2002}. Consequently, the peaked core density profile of closed geometries could allow for core plasma densities exceeding the Greenwald density \cite{MartinGreenwald_2002}. However, density gradients in the pedestal region can also strongly impact the plasma current profile and adversely impact the transport of main ions and impurities (through trapped electron modes, for example) \cite{angioni2017density,skyman2012impurity,ernst2016role}.

Due to the consistent differences in hydrogenic transport, the rate of neutral pumping in the divertor is also massively enhanced in the closed geometry compared to the open geometry. This conclusion is clear and will likely generalise across different simulations or experiments. However, for tokamak reactors helium pumping is a significant concern, and one should not unequivocally conclude from the enhanced hydrogenic pumping that helium pumping would be similarly affected. To estimate this, the full transport of helium must be considered, particularly if a significant amount of helium is directed towards the inner divertor. Enhanced pumping is a clear benefit of the closed divertor, not necessarily because all reactors will desire very high pumping rates, but rather because intrinsically higher pumping in a geometry reduces the engineering requirements for the pump system.

Another conclusion that seems very consistent in this work is that open divertors show power loss along the entire SOL, and significant radiation in the main chamber. This is a direct consequence of the profiles of hydrogenic density and pressure, is consistent at all power levels, and the differences here likely generalise. However, the radiation profile in a real reactor will be strongly impacted by the transport of injected impurities. Strongly baffled divertors can impact the impurity enrichment in the divertor, though even this is not clear-cut \cite{Sun_2023,pitcher2000effect,Casali_2022}. After all, divertors with strong baffling have stronger neutral compression, but they also have weaker plasma flows upstream, which could raise impurity leakage. Hence, it is likely the open divertors will generally have more radiation upstream, but this should not be taken as unequivocal fact. 

In a similar vain, the open divertor shows radiation along the SOL moving radially into the core as the plasma is pushed into deep detachment. Again, this is likely to hold true across other studies, though of course this picture may be strongly impacted by impurity transport. If this conclusion is generally a feature of more open divertors, then a machine which needs to access stable control of a detachment front near the target will likely need a closed divertor. It is likely, however, that in very deep detachment, when the volume recombination front moves past the divertor region, a closed and open divertor will behave similarly (though this is still an open question). Therefore, scenarios such as the x-point radiator may see little benefit from operating with a closed divertor throat \cite{lunt2023compact,stroth2022model}.

When it comes to detachment access, it is consistently found that the detachment threshold is lower in the closed geometry. It is important to stress that though detachment in these simulations is accessed by raising the upstream electron density, the important conclusion here does not concern density, but detachment access in general. A low detachment threshold in terms of density likely means that for the same density, a lower impurity seeding rate is needed, or the divertor could operate at higher powers whilst still maintaining detachment \cite{lipschultz2016sensitivity,Cowley_2022}. 
It is also important to note that the difference in detachment access between the two geometries is drastically reduced at higher powers. What’s more, when there is a large difference in detachment access, the closed geometry shows intrinsically higher target loads for a given target temperature. Hence it is not clear whether this lower detachment threshold is extremely advantageous to start with. 

When discussing detachment access and baffling it is important to contextualise this work with previous work. In particular, previous work has consistently found lower detachment thresholds for divertors 
 with stronger baffling \cite{reimerdes2021initial,Casali_2022}. What's more, several studies have noted a higher detachment threshold than expected in geometries with high total flux expansion \cite{fil2020separating,carpita2023reduction}. One study in particular \cite{carpita2023reduction} attributed a higher detachment threshold in TCV to the reduction in pressure caused by strong parallel flows. The work presented here is consistent with this picture, since the presence of strong convective flux leads to reductions in total pressure and electron static pressure in the open divertor; and a corresponding increase in detachment threshold.

Finally, this study has compared the performance of two geometries: an extremely open and closed divertor chamber. However, there are many more subtle changes to plasma facing components and baffling that can modify plasma performance. For example, what has not been considered here is whether the performance characteristics of a closed divertor and tightly baffled divertor differ, and whether the location of baffle throat impacts the plasma performance.

\section{Conclusions}

In this work the impacts of divertor baffling on a core-fuelled hydrogenic plasma have been isolated and studied using SOLPS-ITER simulations. Simulations of a closed divertor shows two orders of magnitude stronger neutral compression and an order of magnitude higher pumping rates than the open divertor. As a result, strong baffling produces core profiles with more peaked core densities than the open divertor when fuelled from the core. For reactors which have difficult pumping requirements, or devices which aim to have peaked density profiles driven by pellet fuelling, a closed divertor is advantageous.

As a result of the different neutral transport and plasma fuelling, the two geometries show significantly different radiation profiles. In the closed geometry, most of the plasma power loss occurs in the divertor region localised near the target at the threshold of detachment. This is in contrast to the open geometry, which shows gradual radiation along the entire SOL, including in the main chamber. As detachment evolves, this radiation profile rapidly moves further upstream and inwards towards the core, quickly dropping upstream temperatures across the entire SOL. On the other hand, the closed divertor shows slow and stable poloidal movement of a localised detachment front near the target.

Comparing simulations of a closed and open MAST-U geometry, the closed geometry shows easier access to detachment, and can operate with a significantly lower midplane collisionality at the threshold of detachment. However, the closed geometry also has slightly higher target fluxes when attached. This is mainly because of the higher upstream temperatures and lower pressure losses along the `killer' flux tube in the closed divertor. The pressure losses in the open divertor are enhanced by the presence of strong upstream flows due to main chamber ionisation. As power is increased, the static pressure and detachment access characteristics in the `killer' flux tube becomes more similar between the two geometries. This calls into question whether benefits of baffling measured on present-day machines may extrapolate to reactor-level devices.

\vspace{1cm}

\large
\textbf{References}

\normalsize
\vspace{1cm}

\bibliography{references.bib} 
\bibliographystyle{vancouver}

\appendix

\section{Artificial Nitrogen Impurity}\label{appendix}

\begin{figure}[th]
    \centering
    \includegraphics[width=0.7\linewidth]{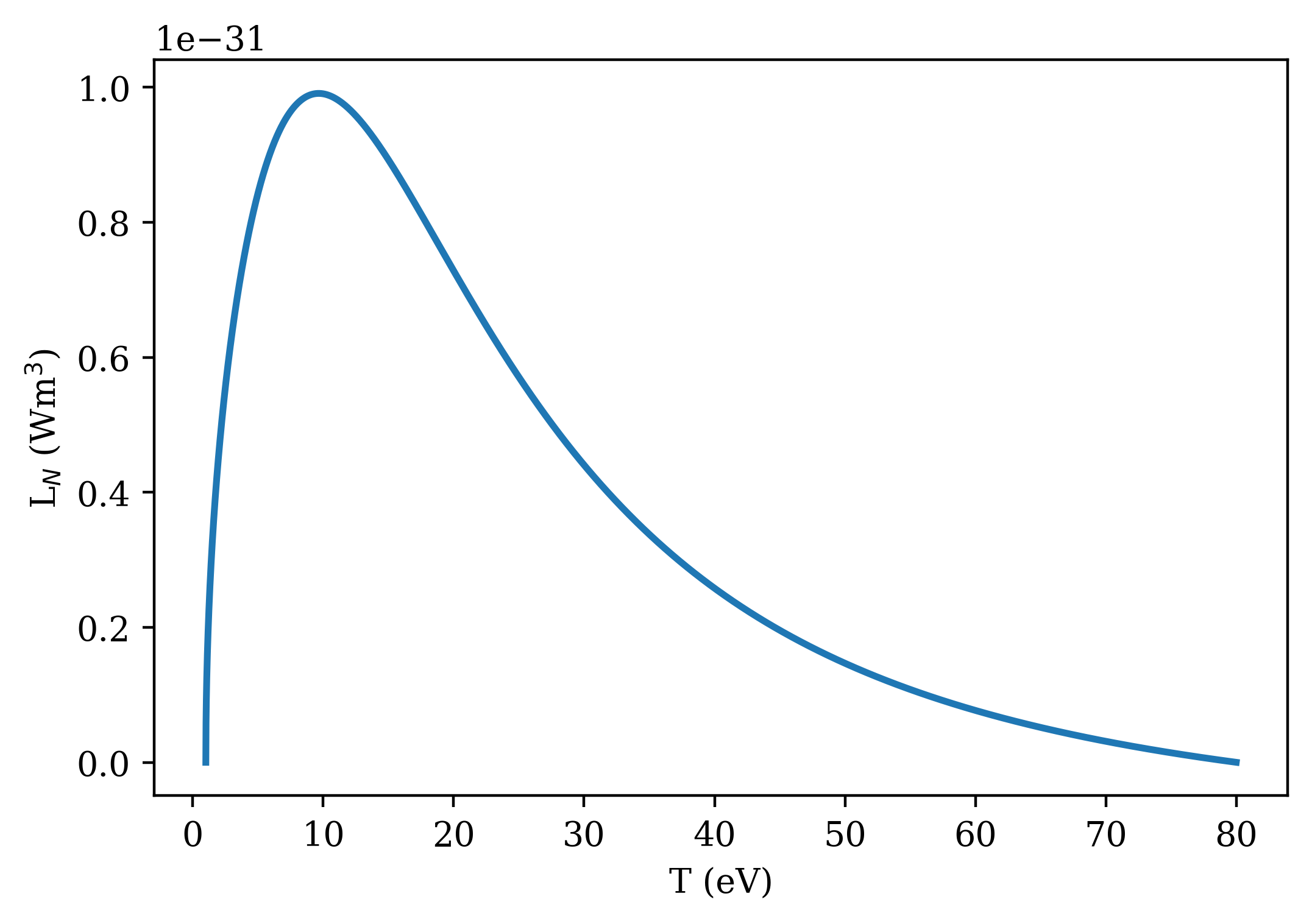}
    \caption{An analytically approximated electron cooling function for Nitrogen, the form for which is given in Equation \ref{coolingfuncEq}.}
    \label{fig:nitrogencoolingFunc}
\end{figure}

The radiation model used for this study was an artificial fixed-fraction nitrogen impurity, with the following analytic cooling curve:

\begin{equation}
  L_{N}(T)=
    5.9\times10^{-34} \frac{\left(T-1\right)^{1/2}\left(80-T\right)}{1+3.1\times10^{-3} \left(T-1 \right)^{2}},
\label{coolingfuncEq}
\end{equation}

valid between 1 and 80 eV and equal to 0 otherwise. This artificial cooling curve is shown in Figure \ref{fig:nitrogencoolingFunc}.

\end{document}